\documentclass[final,3p,times,twocolumn]{elsarticle}

\usepackage{amssymb}
\usepackage{lineno}
\usepackage[figuresright]{rotating}
\usepackage[cmex10]{amsmath}
\usepackage[figuresright]{rotating}
\usepackage[linesnumbered, ruled]{algorithm2e}
\usepackage{subfig}
\usepackage{url}
\usepackage{hyperref}
\usepackage{inputenc}
\usepackage{enumitem}

\begin{document}

\begin{frontmatter}

%\dochead{}

\title{The GPU vs Phi Debate:\\Risk Analytics Using Many-Core Computing\tnoteref{t1}\tnoteref{t2}}

\tnotetext[t1]{A modified version of this article is accepted to the Computers \& Electrical Engineering Journal under the title - \textit{``The Hardware Accelerator Debate: A Financial Risk Case Study Using Many-Core Computing''} DOI: 10.1016/j.compeleceng.2015.01.012.}
\tnotetext[t2]{Please cite the article as - \textit{Blesson Varghese, ``The Hardware Accelerator Debate: A Financial Risk Case Study Using Many-Core Computing,'' Computers \& Electrical Engineering, 2015}.}

\author[sta]{Blesson Varghese}
\ead{varghese@st-andrews.ac.uk}
\ead[url]{http://www.blessonv.com}

\address[sta]{School of Computer Science, University of St Andrews, UK}

\begin{abstract}
The risk of reinsurance portfolios covering globally occurring natural catastrophes, such as earthquakes and hurricanes, is quantified by employing simulations. These simulations are computationally intensive and require large amounts of data to be processed. The use of many-core hardware accelerators, such as the Intel Xeon Phi and the NVIDIA Graphics Processing Unit (GPU), are desirable for achieving high-performance risk analytics. In this paper, we set out to investigate how accelerators can be employed in risk analytics, focusing on developing parallel algorithms for Aggregate Risk Analysis, a simulation which computes the Probable Maximum Loss of a portfolio taking both primary and secondary uncertainties into account. The key result is that both hardware accelerators are useful in different contexts; without taking data transfer times into account the Phi had lowest execution times when used independently and the GPU along with a host in a hybrid platform yielded best performance.
\end{abstract}


\end{frontmatter}

%\linenumbers
%% main text

\section{Introduction}
Risk analytics \cite{riskanalytics-1} has become an integral part of a business process in a range of domains (for example, \cite{catevent-1,catevent-2,structrisk-1,chemrisk-1,disease-1,nuclear-1,radio-1}). Large datasets are consumed by a risk model and hundreds of thousands or even millions of time consuming simulations are performed. Here the application of parallel and high-performance computing techniques are attractive. 

Interestingly, in the financial risk domain, specifically insurance and reinsurance, where data sizes are as large or even larger than what is employed in the above domains, relatively fewer parallel and high-performance computing techniques have been applied. Given the dependencies of the insurance and reinsurance setting on volatile markets, simulations that can be performed in a timely manner are essential. 

\subsection{Background}
Companies hold portfolios of contracts that cover risks associated with catastrophic events such as earthquakes, hurricanes and floods. In order to have a marketplace for such risk it is critical to be able to efficiently quantify individual risks and portfolios of risks. The analytical pipeline of the modern quantitative insurance or reinsurance company typically consists of three major stages, namely risk assessment \cite{ra1,ra2}, portfolio risk management and pricing \cite{prm1}, and enterprise risk management \cite{erm1}.

In the first stage, catastrophe models \cite{catmodel-1} are used to provide loss estimates by taking two inputs. Firstly, stochastic event catalogues which are a mathematical representation of the natural occurrence patterns and characteristics of catastrophes. Secondly, exposure databases that describe thousands or millions of buildings to be analysed, their construction types, location, value, use, and coverage. Each event-exposure pair is then analysed by a risk model that quantifies the hazard intensity at the exposure site, the vulnerability of the building and resulting damage level, and the expected loss, given the customer's financial terms. The output of a catastrophe model is an Event Loss Table (ELT) which specifies the probability of occurrence and the expected loss for every event in the catalogue. 

In the second stage of the analysis pipeline, portfolio risk management and pricing of portfolios of contracts necessitates a further level of stochastic simulation, called Aggregate Risk Analysis or referred to as ARA in this paper \cite{s1,s2,s3,s4,s5}. ARA is a Monte Carlo like simulation in which each trial represents an alternative view of which catastrophic events occur and in which order they occur within a predetermined period or a contractual year. In order to provide actuaries and decision makers with a consistent lens through which to view results, a pre-simulated Year Event Table (YET) containing a million alternative views of a single contractual year is employed. The output of ARA is a Year Loss Table, in which the results are highly aggregated. 

From the output of ARA, a reinsurer can derive important portfolio risk metrics such as the Probable Maximum Loss (PML) \cite{pml1,pml2} and the Tail Value-at-Risk (TVaR) \cite{tvar1,tvar2}. The output is then interpreted by actuaries for key internal decision making, planning a financial year and reporting to regulators and rating agencies. Furthermore, these metrics then flow into the final stage in the risk analysis pipeline, namely Enterprise Risk Management, where liability, asset, and other forms of risks are combined and correlated to generate an enterprise wide view of risk.

\subsection{Problems}
There are two problems that need to be addressed for achieving high-performance risk analytics. Both problems can be solved if the data, memory and computational challenges of the analysis can be efficiently addressed. The first problem is related to developing methods for applying uncertainties. ARA presented above accounts for only `Primary Uncertainty', which is the uncertainty associated with whether an event occurs or not in a simulated year. However, there is `Secondary Uncertainty', which captures the uncertainty in the level of loss due to the use of simplified physical models and limitations in the available data. 

There are many sources of this uncertainty that need to be taken into account when considering catastrophic risk, including unknown exposure and hazard parameters and their interaction. For example, the exposure data which describes the buildings, their locations, and construction types may be incomplete, lacking in sufficient detail, or may be simply inaccurate. Also the physical modelling of hazard, for example an earthquake, may naturally generate a distribution of hazard intensity values due to uncertainty associated with the energy attenuation functions used or driving data such as soil type. Lastly, building vulnerability functions are simplifications of complex physical phenomenon and are therefore much better at producing loss distributions than accurate point estimates. Hence, there is a need to develop methods to not only capture primary uncertainty but also quantify secondary uncertainty in risk analysis.

The analysis uses mean loss values when only primary uncertainty is accounted for. Using such discrete values is an oversimplification, because in reality for any event there is a multitude of possible loss outcomes resulting in a distribution of potential losses. A simulation taking a distribution of losses requires statistical tools, for example, the beta probability distribution to estimate the loss using inverse beta cumulative density function which are both data intensive and computationally intensive. Such an analysis will need to accept as input complete event loss distributions represented by the event rate, mean loss, independent standard deviation, and correlated standard deviation, and better account for the range of possible outcomes.

The second problem is related to implementing parallel risk analysis methods for achieving timely results. 
From a computational perspective the ARA simulation differs from other Monte Carlo simulations since trials are pre-simulated, rather than randomly generated on-the-fly. 
This provides millions of alternate views of a contractual year comprising thousands of events which are pre-simulated as a YET. 
From an analytical perspective, a pre-simulated YET lends itself to statistical validation and to tuning for seasonality and cluster effects. 
However, there are significant challenges in achieving efficient parallelisation. 
The extremely large YET must be carefully shared between processing cores if the computation is to achieve good speed-up when there is limited memory bandwidth.

\subsubsection{Addressing the problems}
In this paper, we investigate hardware acceleration platforms for ARA. Parallel algorithms for ARA that initially take primary uncertainty into account is implemented. Further, a methodology that considers secondary uncertainty is presented. The algorithms are implemented on the Intel Phi Coprocessor and an NVIDIA Graphics Processing Unit (GPU). Experimental studies evaluate how ARA performs on the hardware accelerators independently and along with a host processor. 

Both the Phi Coprocessor and the GPU are competing hardware accelerators that offer alternative machine architectures to that of a regular CPU. 
While both hardware accelerators are significantly different, they provide in common, firstly, lots of cycles for independent parallelism, secondly, fast memory access under the right circumstances, and finally, fast mathematical computations. 
The parallel algorithms implemented in the paper take full advantage of the high levels of parallelism, fast shared memory access and fast numerical performance. In this research, the algorithm exploits the machine architecture of the accelerators to achieve parallelism and fast memory access and for performing fast numerical computations. The result is a speed up of 16x - 21x that is achieved on the Phi and GPU over respective baseline sequential implementations on the CPU. The experiments show the feasibility of hardware accelerators offering a relatively low cost high-performance computing solution for performing fast ARA. 

\subsection{Related Work}
The domain of computational finance and risk addresses problems related to achieving fast computations, surmounting challenges of data management and efficiently handling memory of computing systems. Parallelism is exploited to achieve this (for example, \cite{finpar-1,finpar-2,finpar-3,finpar-4}). Therefore, this domain is dependent on the advances in high-performance computing. Research on financial applications for production-based computing systems have progressed from small scale clusters \cite{smallcluster1,smallcluster2} to large supercomputers \cite{supercomputer1,supercomputer2}. These applications are hosted either on in-house clusters or on supercomputers. 

A number of financial applications are migrated from small clusters to multi-core and many-core processors which are available at a low budget \cite{budgetplatform1}. For example, research related to financial applications exploiting the Cell BE processor is reported in \cite{cellbe-1,cellbe-2}. FPGAs are another alternative platform presented in \cite{fpga1,fpga2,fpga3,fpga4}. GPU acceleration is employed more recently \cite{gpu1,gpu2,gpu3,gpu5}. Heterogeneous clusters comprising hardware accelerators are now employed \cite{hetcluster-1,hetcluster-2}. In all the above research, the need for speeding up financial applications are presented and is achieved. There is limited research on how high-performance computing advances can be applied to simulations in the risk domain, specifically the insurance and reinsurance setting, which is considered in this paper. 

Although high-performance computing platforms are an option to accommodate and accelerate risk simulations, there is an investment cost that will need to be borne along with maintenance costs. A relatively cost effective solution is to employ hardware accelerators to address the problems faced by current risk simulations. Hardware accelerators can provide fast numerical computations which are required by statistical functions that support applying secondary uncertainty in ARA. Additionally, accelerators can be exploited to achieve speed up and thereby use risk simulations in real-time. 

The remainder of this paper is organised as follows. Section \ref{aggregateriskanalysis} proposes an algorithm for performing ARA with primary uncertainty. Section \ref{secondaryuncertainty} presents a methodology for applying secondary uncertainty and the inputs required for the extended analysis. Section \ref{implementation} presents the implementations of algorithms both for incorporating primary and secondary uncertainties on multi-core and many-core hardware platforms. Section \ref{results} considers the results obtained from the implementations in the experimental studies. Section \ref{conclusions} concludes this paper.
\label{introduction}

\section{Aggregate Risk Analysis}
\label{aggregateriskanalysis}
\begin{figure*}
	\centering
	\includegraphics[width = 0.63\textwidth]{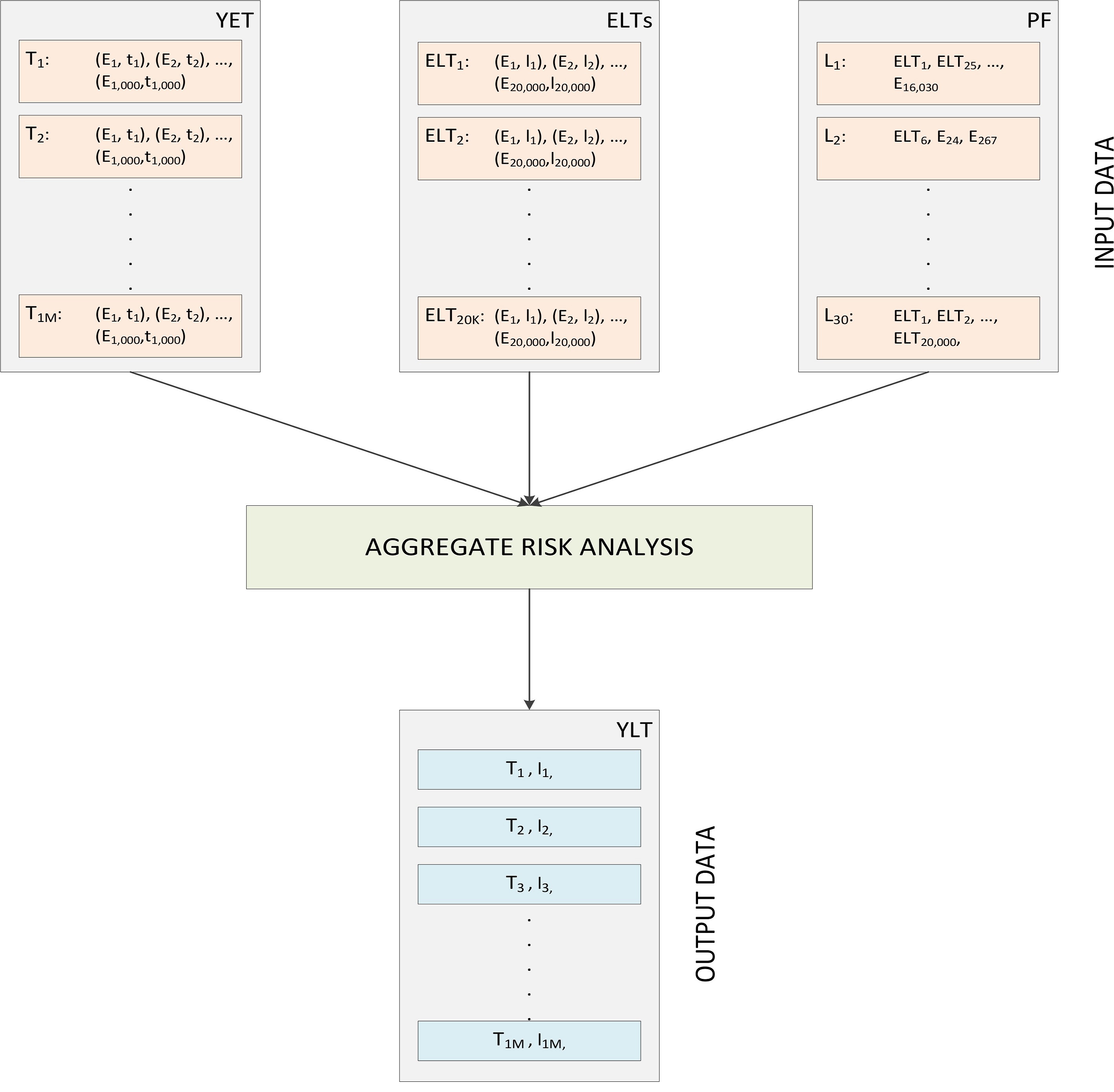}
	\caption{Inputs and outputs of Aggregate Risk Analysis}
	\label{figure0}
\end{figure*}

Stochastic Monte Carlo like simulations are required for portfolio risk management and contract pricing, and Aggregate Risk Analysis (ARA) is one such simulation. The merit of performing the analysis is that millions of alternative views of a single contractual year can be obtained. This section will consider the inputs required for performing ARA, propose an algorithm for ARA, present the financial terms employed in the algorithm, and highlight the output of the analysis (refer Figure \ref{figure0}). 

\subsection{Inputs}
The following three data tables are required for performing ARA:

\begin{enumerate}[leftmargin=12pt]
\item[i.] The Year Event Table (YET), denoted as $YET$, is a database of pre-simulated occurrences of events from a catalogue of stochastic events. Each record in a YET called a `trial', denoted as $T_i$, represents a possible sequence of event occurrences for any given year. The sequence of events is defined by an ordered set of tuples containing the ID of an event and the time-stamp of its occurrence in that trial represented as $T_i = \{(E_{i, 1}, t_{i, 1}), \dots, (E_{i, k}, t_{i, k})\}$.

The set is ordered by ascending time-stamp values. A typical YET may comprise thousands to millions of trials. Each trial may have approximately between 800 to 1500 `event time-stamp' pairs based on a global event catalogue covering multiple perils. The YET can be represented as 
\begin{center}
$YET	= \{ T_i = \{(E_{i, 1}, t_{i, 1}), \dots, (E_{i, k}, t_{i, k})\} \}$,\\ 
where $i = 1, 2, \dots$ and $k = 1, 2, \dots, 1500$.
\end{center}

\item[ii.] Event Loss Tables, denoted as $ELT$, represent collections of specific events and their corresponding losses with respect to an exposure set. Each record in an ELT is denoted as $EL_{i} =	\{E_{i}, l_{i}\}$ and the financial terms associated with the ELT are represented as a tuple $\mathcal{I} = (\mathcal{I}_{1}, \mathcal{I}_{2}, \dots)$. 

A typical aggregate analysis may comprise 10,000 ELTs, each containing 10,000 to 30,000 event losses with exceptions of even up to 2,000,000 event losses. The ELTs can be represented as 
\begin{center}
$
ELT=\left\{
	\begin{array}{l c l}
	EL_{i}			&	=	&	\{E_{i}, l_{i}\},\\
	\mathcal{I} 		&	=	&	(\mathcal{I}_{1}, \mathcal{I}_{2}, \dots)
	\end{array}\right\}$, \\
with $i = 1, 2, \dots , 30,000$.
\end{center}

\item[iii.] A Portfolio, denoted as $PF$ contains a group of Programs, denoted as $P$ represented as $PF = \{P_{1}, P_{2}, \cdots, P_{n}\}$, with $n = 1, 2, \dots, 10$.

Each Program in turn covers a set of Layers, denoted as $L$, which covers a collection of ELTs under a set of layer terms. A single layer is composed of two attributes. Firstly, the set of ELTs $\mathcal{E} = \{ELT_1, ELT_2, \dots, ELT_j\}$, and secondly, the Layer Terms, denoted as $\mathcal{T} = (\mathcal{T}_{OccR}, \mathcal{T}_{OccL}, \mathcal{T}_{AggR}, \mathcal{T}_{AggL})$. The Layer occurrence terms are (i) Occurrence Retention, denoted as $\mathcal{T}_{OccR}$, which is the retention or deductible of the insured for an individual occurrence loss, and (ii) Occurrence Limit, denoted as $\mathcal{T}_{OccL}$, which is the limit or coverage the insurer will pay for occurrence losses in excess of retention. The Layer aggregate terms are (i) Aggregate Retention, denoted as $\mathcal{T}_{AggR}$, which is the retention or deductible of the insured for an annual cumulative loss, and (ii) Aggregate Limit, denoted as $\mathcal{T}_{AggL}$, which is the limit or coverage the insurer will pay for annual cumulative losses in excess of aggregate retention.

A typical layer covers approximately 3 to 30 individual ELTs. The Layer can be represented as 

\begin{center}
$L=\left\{
	\begin{array}{l c l}
	\mathcal{E}	& =	& \{ELT_1, ELT_2, \dots, ELT_j\}, \\
	\mathcal{T}	& = & (\mathcal{T}_{OccR}, \mathcal{T}_{OccL}, \mathcal{T}_{AggR}, \mathcal{T}_{AggL})
	\end{array}\right\}$, \\
with $j = 1, 2, \dots, 30$.
\end{center}
	
\end{enumerate}

\subsection{Algorithm}

Algorithm \ref{algorithm1} presents two stages for performing aggregate analysis with primary uncertainty. In the first stage, data is loaded into local memory, referred to as the pre-processing stage. In this stage the $YET$, $ELT$ and $PF$, are loaded into memory.
\begin{algorithm} %[H]
\caption{Aggregate Risk Analysis with Primary Uncertainty}
\label{algorithm1}
\SetAlgoLined
\DontPrintSemicolon

\SetKwInOut{Input}{Input}
\SetKwInOut{Output}{Output}

\BlankLine

\Input{$YET$, $ELT$, $PF$}
\Output{$YLT$}

\BlankLine

\For{each Program, $P$, in $PF$}{
	\For{each Layer, $L$, in $P$}{
		\For{each Trial, $T$, in $YET$}{
			\For{each Event, $E$, in $T$}{
				\For{each $ELT$ covered by $L$}{
					Lookup $E$ in the $ELT$ and find corresponding loss, $l_{E}$\;
%					Apply Secondary Uncertainty to $l_{E}$\;
					$l_{E} \leftarrow$ Apply Financial Terms$(l_{E})$\;
					$l_{T} \leftarrow$ $l_{T}$ + $l_{E}$\;
				}
				$l_{T} \leftarrow$ Apply Occurrence Financial Terms $(l_{T})$\;	
				$l_{T} \leftarrow$ Apply Aggregate Financial Terms $(l_{T})$\;
			}
		}
	}
}			
Populate $YLT$ using $l_{T}$\;
\BlankLine
\end{algorithm}

In the second stage, the four step simulation for each Layer and for each Trial in the YET is performed and the Year Loss Table ($YLT$) is produced. 

In the first step shown in line no. 6, each event of a trial and the corresponding event loss in the set of ELTs associated with the Layer are determined. In the second step shown in line no. 7, a set of contractual financial terms are applied to each loss value of the Event-Loss pair extracted from an ELT to the benefit of the layer. For this the losses for a specific event's set of financial terms $\mathcal{I}$ are accumulated across all ELTs into a single event loss shown in line no. 8. 

In the third step (line no. 10), the event loss for each event occurrence in the trial, combined across all ELTs associated with the layer, is subject to occurrence terms. Occurrence terms are applicable to individual event occurrences independent of any other occurrences in the trial. The occurrence terms capture specific contractual properties of 'eXcess of Loss' \cite{excessofloss-1, excessofloss-2} treaties as they apply to individual event occurrences only. The event losses net of occurrence terms are then accumulated into a single aggregate loss for the given trial. The occurrence terms are applied as $l_{T} = min ( max ( l_{T} - \mathcal{T}_{OccR}, 0), \mathcal{T}_{OccL})$.

In the fourth step (line no. 11), aggregate terms are applied to a trial's aggregated loss of a layer. Unlike occurrence terms, aggregate terms are applied to the cumulative sum of occurrence losses within a trial, and thus, the result depends on the sequence of prior events in the trial. This behaviour captures contractual properties as they apply to multiple event occurrences. The aggregate loss net of the aggregate terms is referred to as the trial loss or the year loss. The aggregate terms are applied as $l_{T} = min ( max ( l_{T} - \mathcal{T}_{AggR}, 0 ), \mathcal{T}_{AggL})$.

\subsection{Output}
The output of ARA is a loss value associated with each trial of the YET. A reinsurer can derive important portfolio risk metrics such as the Probable Maximum Loss (PML) and the Tail Value-at-Risk (TVaR) which are used for both internal risk management and reporting to regulators and rating agencies. Furthermore, these metrics are used in Enterprise Risk Management, where liability, asset, and other forms of risks are combined and correlated to generate an enterprise wide view of risk.

Additional functions can be used to generate reports that will aid actuaries and decision makers. For example, reports presenting Return Period Losses (RPL), Region/Peril losses, Multi-Marginal Analysis and Stochastic Exceedance Probability (STEP) Analysis.

\section{Applying Secondary Uncertainty}
\label{secondaryuncertainty}
The methodology to compute secondary uncertainty builds on industry wide practices. The inputs required for applying secondary uncertainty and the sequence of steps are considered in this section. 

\subsection{Inputs}
For performing ARA accounting for secondary uncertainty requires the following additional inputs: 

\begin{enumerate}[leftmargin=12pt]
\item[i.] $z_{(Prog,E)} = P_{(Prog,E)} \in U(0,1)$, referred to as the Program-and-Event-Occurrence-Specific random number. Each Event occurrence across different Programs have different random numbers. 
\item[ii.] $z_{(E)} = P_{(E)} \in U(0,1)$, referred to as the Event-Occurrence-Specific random number. Each Event occurrence across different Programs have the same random number.
\item[iii.] $\mu_{l}$, referred to as the mean loss is the loss associated with an event $E_{i}$.
\item[iv.] $\sigma_{I}$, referred to as the independent standard deviation of loss, represents the variance within the event-loss distribution.
\item[v.] $\sigma_{C}$, referred to as the correlated standard deviation of loss, represents the error of the event-occurrence dependencies. 
\item[vi.] $max_{l}$, referred to as the maximum expected loss.
\end{enumerate}

To capture the above additional inputs required for computing secondary uncertainty, the inputs used in ARA with primary uncertainty need to be modified. The Year Event Table (YET), denoted as $YET$ is redefined as 

\begin{center}
$YET = \{ T_i = \{(E_{i, 1}, t_{i, 1}, z_{(Prog, E)_{i, 1}}), \dots,\newline (E_{i, k}, t_{i, k}, z_{(Prog, E)_{i, k}})\} \}$,\\
where $i = 1, 2, \dots$ and $k = 1, 2, \dots, 1500$.
\end{center}

Extended Event Loss Tables, denoted as $XELT$, are used instead of $ELT$. Each record in an XELT is denoted as `eXtended' Event Loss $XEL_{i} = \{E_{i}, l_{i}, z_{(E)_{i}}, \sigma_{I_{i}}, \sigma_{C_{i}}, max_{l_{i}}\}$
and the financial terms associated with the XELT are represented as a tuple $\mathcal{I} = (\mathcal{I}_{1}, \mathcal{I}_{2}, \dots)$.

The extended ELT is represented as $XELT=$
\begin{center}
$\left\{
	\begin{array}{l c l}
	XEL_{i}		&	=	&	\{E_{i}, \mu_{l_{i}}, z_{(E)_{i}}, s\sigma_{I_{i}}, \sigma_{C_{i}}, max_{l_{i}} \},\\
	\mathcal{I} 	&	=	&	(\mathcal{I}_{1}, \mathcal{I}_{2}, \dots)
	\end{array} \right\}$, \\
	with $i = 1, 2, \dots , 30,000$.
\end{center}

The representation of the Layer is modified as $L=$
\begin{center}
$\left\{
	\begin{array}{l c l}
	\mathcal{E}	& =	& \{XELT_1, XELT_2, \dots, XELT_j\}, \\
	\mathcal{T}	& = & (\mathcal{T}_{OccR}, \mathcal{T}_{OccL}, \mathcal{T}_{AggR}, \mathcal{T}_{AggL})
	\end{array}\right\}$,\\
	with $j = 1, 2, \dots, 30$.
\end{center}

\subsection{Combining standard deviation}

Given the above inputs, the independent and correlated standard deviations need to be combined to reduce the error for estimating the loss value associated with an event. This is done in the following five steps: 

\begin{enumerate}[leftmargin=12pt]
\item[i.] The raw standard deviation is produced as 
\begin{center}
$\sigma = \sigma_{I} + \sigma_{C}$. 
\end{center}

\item[ii.] The probabilities of occurrences, $z_{(Prog,E)}$ and $z_{(E)}$ are transformed from uniform distribution to normal distribution using $f(x; \mu, \sigma^{2}) = \int\limits_{-\infty}^{x} \frac{1}{\sigma\sqrt{2\pi}}e^{-\frac{1}{2}\big({\frac{x - \mu}{\sigma}}\big)^{2}}\mathrm{d}x$.

This is applied to the probabilities of event occurrences as 
\begin{center}
$v_{(Prog,E)} = f(z_{(Prog,E)}; 0, 1) \in N(0,1)$ and 
$v_{(E)} = f(z_{(E)}; 0, 1) \in N(0,1)$.
\end{center}

\item[iii.] The linear combination of the transformed probabilities of event occurrences and the standard deviations is computed as 
\begin{center}
$LC = v_{(Prog,E)}\Big(\frac{\sigma_{I}}{\sigma}\Big) + v_{(E)}\Big(\frac{\sigma_{C}}{\sigma}\Big)$.
\end{center}

\item[iv.] The normal random variable is computed as 
\begin{center}
$v = \frac{LC}{\sqrt{{\big(\frac{\sigma_{I}}{\sigma}\big)}^2 + {\big(\frac{\sigma_{C}}{\sigma}\big)}^2}}$.
\end{center}

\item[v.] The normal random variable is transformed from normal distribution to uniform distribution as 
\begin{center}
$z = \Phi(v) = F_{Norm}(v) = \frac{1}{\sqrt{2\pi}}\int\limits_{-\infty}^{v}e^{\frac{-t^2}{2}}\mathrm{d}t$.
\end{center}

\end{enumerate}

The model used above for combining the independent and correlated standard deviations represents two extreme cases. The first case in which $\sigma_{I} = 0$ and the second case in which $\sigma_{C} = 0$. The model also ensures that the final random number, $z$, is based on both the independent and correlated standard deviations.

\subsection{Estimating Losses}

The loss is estimated using the Beta distribution since fitting such a distribution allows the representation of risks quite accurately. The Beta distribution is a two parameter distribution, with an upper bound for the standard deviation. The standard deviation is $\sigma_{\beta} = \frac{\sigma}{max_{l}}$ and the mean is $\mu_{\beta} = \frac{\mu_{l}}{max_{l}}$. The alpha value is 
\begin{center}
$\alpha = \mu_{\beta}\Big(\big({\frac{\sigma_{\beta_{max}}}{\sigma_{\beta}}\big)}^{2}-1\Big)$,
\end{center}
and the beta value is 
\begin{center}
$\beta = (1 - \mu_{\beta})\Big(\big({\frac{\sigma_{\beta_{max}}}{\sigma_{\beta}}\big)}^{2}-1\Big)$
\end{center}

An upper bound is set to limit the standard deviation using $\sigma_{\beta_{max}} = \sqrt{\mu_{\beta}(1-\mu_{\beta})}$, if $\sigma_{\beta} > \sigma_{\beta_{max}}$, then $\sigma_{\beta} = \sigma_{\beta_{max}}$. In the algorithm reported in this paper, for numerical purposes, a value very close to $\sigma_{\beta_{max}}$ is chosen.

The loss after applying beta distribution functions are obtained as 
\begin{center}
$Loss = max_{l} * InvCDF_{beta}(z; \alpha, \beta)$, \\
$InvCDF_{beta}(z; \alpha, \beta) = \Big(\frac{B(z; \alpha, \beta)}{B(\alpha, \beta)}\Big)^{-1}$, \\
where $B(z; \alpha, \beta) =	\int\limits_{0}^{z} t^{\alpha - 1}(1 - t)^{\beta - 1}\mathrm{d}t$.
\end{center}

Algorithm \ref{algorithm1} requires the redefined inputs for applying secondary uncertainty. 
The modified $YET$, the $XELT$ and the modified Layer, $L$ in $PF$ considered in this section are used as inputs. 
The above steps to compute secondary uncertainty are incorporated after line no. 6 and before applying the financial terms to each event loss. 

\section{Implementation}
\label{implementation}
The hardware platforms on which ARA is implemented are firstly considered in this section, followed by a discussion on the data structures used in the implementation of ARA and on the methods for computing secondary uncertainty. Optimisations incorporated in the implementations are further considered. 

\subsection{Experimental Platforms}
Four hardware platforms ranging from desktop CPUs, such as the Intel i7, a non-consumer workstation/server, such as the Intel Xeon, and accelerators, such as the NVIDIA GPU and the Intel Phi, are used for implementing sequential and parallel ARA algorithms.  

Firstly, a multi-core CPU is employed whose specifications are a 3.40 GHz quad-core Intel(R) Core (TM) i7-2600 processor with 16.0 GB of RAM. The processor has 256 KB L2 cache per core, 8MB L3 cache and a maximum memory bandwidth of 21 GB/sec. 

Secondly, two multi-core CPUs are employed whose specifications are a 2.00 GHz octa-core Intel (R) Xeon(R) E5-2650 processor with 256 GB of RAM. The processor has a 20 MB cache and a maximum memory bandwidth of 51.2 GB/sec. 

Both the above processors support Advanced Vector Extensions (AVX) instructions for vector operations. Both sequential and parallel versions of the ARA algorithm were implemented on these platforms. The sequential version adopted C++, while the parallel version was implemented in C++ and OpenMP. The versions were compiled using the icpc compiler, version 13.1 provided by the Intel(R) Compiler Suite. The `\texttt{\small -O3}' compiler flag was used for optimisation, and the OpenMP directive and the Intel Math Kernel library were included.
%the `\texttt{\small -openmp}' was used to include the OpenMP directive and the `\texttt{\small -mkl}' flag to include the Intel Math Kernel library. 

Thirdly, an NVIDIA Tesla C2075 GPU, consisting of 448 CUDA cores, each with a frequency of 1.15 GHz, a global memory of 6 GB and a memory bandwidth of 144 GB/sec. The peak double precision floating point performance is nearly 0.515 Tflops. The implementation of the algorithm is compiled using the NVIDIA CUDA Compiler (nvcc), version 5.0\footnote{\url{https://developer.nvidia.com/cuda-toolkit}}. The implementation is compiled using `\texttt{\small -arch sm\_13}'.

Fourthly, an Intel Xeon Phi Coprocessors 5110P consisting of 60 cores with a frequency of 1.053 GHz. The coprocessor supports a maximum of 240 threads, and a memory of 8 GB and a memory bandwidth of 320 GB/sec is available. The peak double precision floating point performance is close to one Tflop. The Phi Coprocessor is based on the Intel (R) Many Integrate Cores (MIC) architecture and supports the Intel Initial Many-Core Instructions (IMCI). 

The GPU is employed in two ways. Firstly, in a hybrid mode, which refers to execution both on the host and the accelerator (operates in asynchronous mode and performs side-by-side processing with the host device). Secondly, in a non-hybrid mode, which refers to the host supporting the data pre-processing and transfer activities and the accelerator performs all the computation. The Phi on the other hand is not only employed in the hybrid mode, but also in the native mode, which refers to the execution of the code on the accelerator independent of the host device.

In this paper, sequential and parallel implementations of ARA on all the above four platforms, on the hybrid of i7 and GPU, and on the hybrid of Xeon and Phi are considered. Four statistical libraries, namely the Intel MKL, Boost, Beta\_nc and CUDA Math are explored for applying secondary uncertainty. 

\subsection{Implementing Data Structures for the Algorithm}
In ARA, the losses of events in a trial need to be determined by looking up losses in the XELT. The key design question is whether the data structure containing the event-loss pairs of all trials need to be a sparse matrix in the form of a direct access table or a compact representation. While fast lookups can be obtained in the sparse matrix representation, this performance is achieved at the cost of high memory usage. Consider a YET with 1,000,000 events and one Layer with 16 XELTs, and each XELT consisting of 20,000 events with non-zero losses. The representation using a direct access table would require memory to hold $1,000,000 \times 16 = 16,000,000$ event-loss pairs (without considering the data required for computing secondary uncertainty). While such a large data structure is held in memory, 15,700,000 events represent zero loss value.

The sparse representation requires large amount of memory, but it is chosen over any compact representation for the following reason. A search operation is required to find an event-loss pair even in a compact representation. If sequential search is adopted, then $O(n)$ memory accesses are required to find an event-loss pair. If sorting is performed in a pre-processing phase to facilitate a binary search, then $O(log(n))$ memory accesses are required to find an event-loss pair. If a constant-time space-efficient hashing scheme, such as cuckoo hashing \cite{cuckoohashing} is adopted then an event-loss pair can be accessed with a constant number of memory accesses. However, this can be only be achieved at the expense of a complex implementation and overheads depreciating run-time performance. Further, such an implementation on hardware accelerators with complex memory hierarchies is cumbersome. Although large memory space is required for a direct access table, looking up event-loss pairs can be achieved with fewer memory accesses compared to the memory accesses in a compact representation.

The GPU implementation of the algorithm uses global memory to store all data structures. The parallel implementations on the GPU require high memory transactions. This is surmounted by utilising shared memory wherever possible over global memory, but with challenges in dividing data according to memory access patterns of the threads.

Two data structure implementations of 16 XELTs were considered for the GPU. In the first implementation, each XELT is considered as an independent table; therefore, in a read cycle, each thread independently looks up its events from the XELTs. All threads within a block access the same XELT. 

In the second implementation, the 16 XELTs are combined into a single table. In this case it is challenging to divide data according to the access patterns. This results in threads within a block requiring to access different XELTs and loading many rows of the XELT into memory for each block. There are higher memory accesses and transactions between the global and shared memories so that the threads can access the required events and losses from the XELTs. The implementation with combined XELTs performed poorly compared to the implementation which uses independent XELTs. When only primary uncertainty was considered the execution time increased by nearly 15\% and there is a further increase when secondary uncertainty is taken into account.  

\subsection{Implementing Methods to Compute Secondary Uncertainty}
Three statistical functions are required in the method for applying secondary uncertainty. They are (i) the Cumulative Distribution Function (CDF) of Normal distribution, (ii) the Quantile of the CDF of Normal distribution, and (iii) the Quantile of the Beta distribution. The Quantile of the Beta distribution is a numerically intensive function since an iterative method is required for converging at a solution within a certain error bound. 

Four statistical libraries are used to implement the statistical functions required for applying secondary uncertainty. Firstly, the Intel MKL library is used in the implementations on the CPU and Phi for the CDF of the Normal distribution and the Quantile of the Normal distribution \cite{intelmkl-1}. The Intel MKL API currently does not support Beta distribution functions. 

Secondly, the CUDA Math library\footnote{\url{http://docs.nvidia.com/cuda/cuda-math-api/index.html}} is employed. The CDF of the Normal distribution and the Quantile of the Normal Distribution are fast methods that are included in the implementation. The CUDA Math API currently does not support Beta distribution functions.

Thirdly, the Boost statistical library is offered by the Boost C++ libraries\footnote{\url{http://www.boost.org/}}. However, Boost is currently not supported for the GPU platform. 

Fourthly, BETA\_NC another C++ library that can evaluate the CDF of the Noncentral Beta distribution is employed \cite{betanc}. This library is ported onto the GPU platform and all the functions in the libraries are implemented as \texttt{\small \_\_device\_\_} functions for the GPU. This function is also used for the Phi. 

\subsection{Optimising the Implementation on the CPU}
The optimised implementation on the CPUs and the Phi coprocessor takes advantage of vectorisation by making use of Intel's AVX and Phi's IMCI instructions. Single Instruction Multiple Data (SIMD) instructions facilitate eight (in the case of i7 or Xeon) and sixteen (on the Phi) single-precision floating point operations in the time they take for completing a single floating point operation.

ARA incorporating primary uncertainty does not benefit from vectorisation as most computations are memory intensive. However, in ARA incorporating secondary uncertainty, a large number of floating point arithmetic is performed making it a better candidate to be vectorised. For vectorising the function which applies secondary uncertainty, firstly, several single precision floating point values are modified to single precision arrays. Secondly, instead of applying secondary uncertainty on every Event loss (as in the regular implementations), an array is populated with the Event loss values. Then secondary uncertainty is applied on the array. Such an implementation makes room for performing a number of floating point operations at the same time. 

Functions using the Boost library are replaced with equivalents provided by the Intel MKL library as they are optimised to perform SIMD instructions on arrays. The basic floating point arithmetic in secondary uncertainty was vectorised with the help of the \texttt{\small \#pragma ivdep} directive, which tells the compiler that it is safe to vectorise the arrays.  

Further, to exploit the benefits of vectorising, the inverse CDF of the Beta distribution was implemented incorporating SIMD instructions. The Beta\_nc library was found to lend itself well to vectorisation since it had very few branch conditions, such as break or early return statements. Branch conditions make the use of SIMD instructions more difficult. The inverse CDF of the Beta distribution was vectorised in a similar way to which the function applying secondary uncertainty was vectorised; using an array instead of individual floating point values. 

The function for computing the inverse CDF of the Beta distribution incorporates an algorithm for convergence of the solution. One condition for ensuring convergence is based on the desired accuracy of the result. In some cases a large number of iterations are required for the solution to converge causing the computations to take more time. To optimise the function further, solutions which converge quickly are switched out and replaced with new values whose inverse CDF need to be found. In this way, the time spent on cases requiring large times for convergence can be mitigated by performing convergence in other cases.

\subsection{Optimising the Implementation on Hardware accelerators}

Two approaches were considered for performing ARA with primary and secondary uncertainty on the Intel Phi Coprocessor and the NVIDIA GPU. 

For the Intel Phi coprocessor, in the first approach, referred to as the native approach, the source code for the Xeon platform was built for the Intel Many Integrated Core (MIC) architecture using the Intel compiler's \texttt{\small -mmic} flag. The executable along with dynamic libraries required and all input data were copied onto the coprocessor using the \texttt{\small scp} command. A connection to the Coprocessor was established using \texttt{\small ssh} and ARA was performed the same way as it was on the Xeon platform. Though this approach was easy to adopt, it was not easy to run a 800,000 trial simulation since the time taken for copying the input data onto the coprocessor from the host machine was 10 times more than it did for reading input data on the Xeon platform. 

For the Intel Phi coprocessor a second approach is implemented, referred to as the hybrid approach, in which the host (Xeon processor) is combined with the Phi Coprocessor. In this implementation, the input data is initially loaded into the host memory. The processor then writes a set of trials onto a buffer and sends the buffer to the device (Phi Coprocessor). The Coprocessor then applies secondary uncertainty and aggregates the results as needed in ARA for applying primary and secondary uncertainty. The device then writes back onto a buffer on the host. This implementation can accommodate 800,000 trials unlike the first implementation since limited amount of trials are only handled at once on the device.

The connection from the host to the device is established using Intel's Core Offload Infrastructure (COI) library. A process is created on the device using the library once the source starts executing, and sleeps until the host instructs the COI library to start the device function for applying secondary uncertainty and performing aggregation.

The problem of the Coprocessor memory unable to accommodate the entire simulation in the native approach is surmounted by iteratively sending a chunk of trials through a buffer. Better performance can be achieved in the hybrid approach by having the host and device perform the tasks asynchronously. For example, while the device is applying secondary uncertainty and aggregating a chunk of trials, the host could start chunking the trials for the next buffer, or apply secondary uncertainty and perform aggregation for a different chunk of trials. 

For the GPU implementations, in the hybrid and non-hybrid mode, data from both shared and global memories are migrated to the kernel registry, which has low latency. The numerical computations are made slightly faster by using the CUDA Math APIs support for floating point operations (the compiler flag \texttt{\small -use\_fast\_math} is included).

\section{Experimental Results}
\label{results}

%===========================================================================
%---------------------------------SEQUENTIAL--------------------------------
%===========================================================================
In this section, the results obtained from the experimental studies are presented. Typical industry simulation specifications are considered; from 200,000 to 800,000 trials with each trial comprising 1,000 events, and one Layer covering 16 XELTs (industry practitioners were consulted in choosing the size of the simulation). The combined size of the YET comprising 800,000 trials, the 16 XELTs, and the metadata that defines one Layer is approximately 28 GB for each Layer. So if $N$ Layers are to be used in ARA then the input data will be $N \times 28$ GB. The size of the input data is reduced when the ARA only considers PU since the YET and ELTs are not as large as when SU is considered. We refer to `ARA with PU' when only primary uncertainty is considered and `ARA with SU' when secondary uncertainty is also taken into account. A single library (for example, Boost) or a combination of libraries (for example, MKL and Beta\_nc) is employed for the statistical functions in secondary uncertainty. 

Figure \ref{figure1} and Figure \ref{figure2} are the graphs obtained for the sequential implementation. Figure \ref{figure3} to Figure \ref{figure9} are the results obtained for parallel implementations on the CPUs, on hardware accelerators and on hybrid platforms. The summary of the results from both sequential and parallel implementations are presented in Figure \ref{figure13}.

\subsection{Sequential Implementation}

\begin{figure} %[!tp][t]
\centering
	\subfloat[i7]{\label{fig:1a}\includegraphics[width=0.5\textwidth]{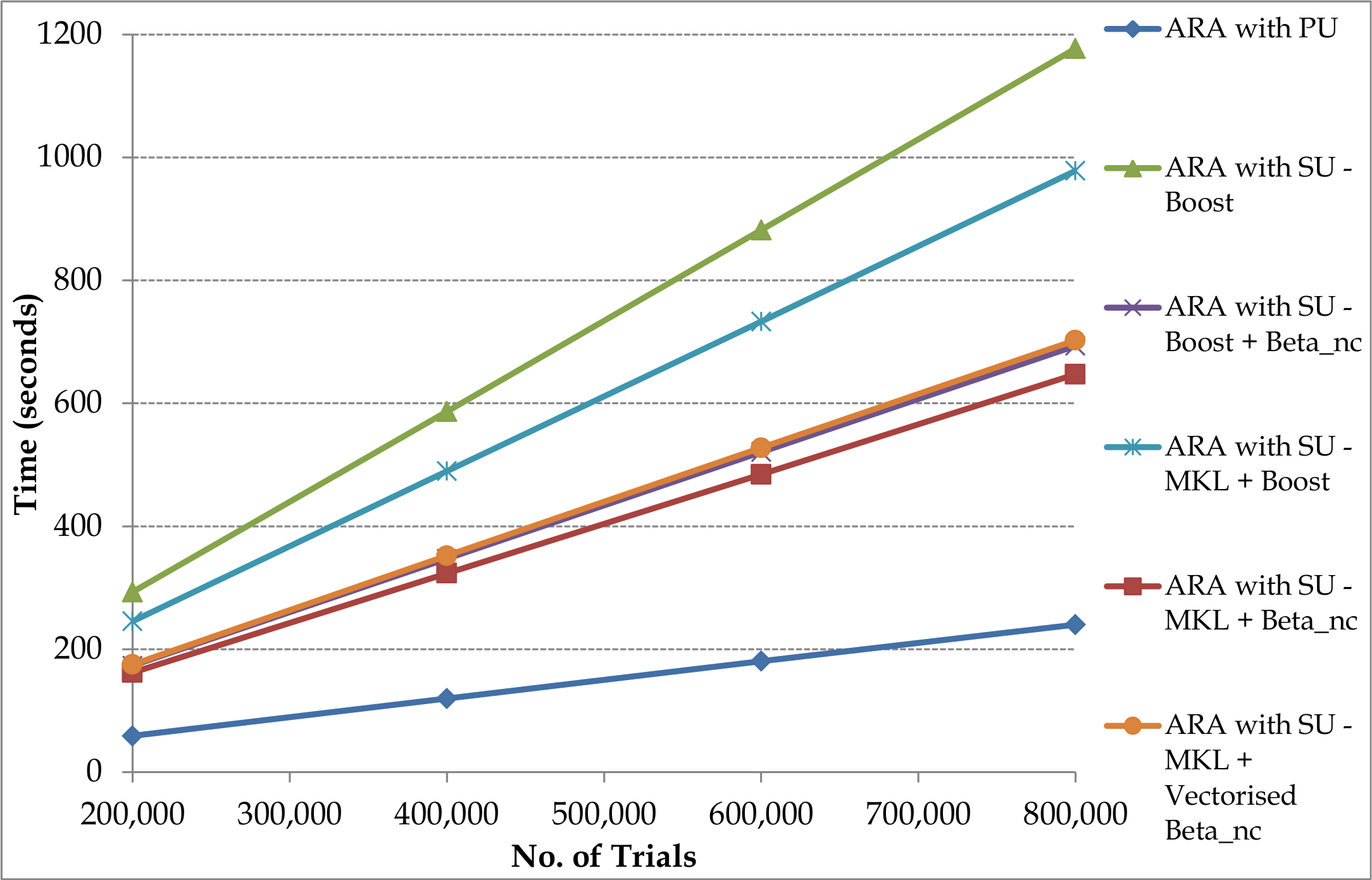}} \\
	\subfloat[Xeon]{\label{fig:1b}\includegraphics[width=0.5\textwidth]{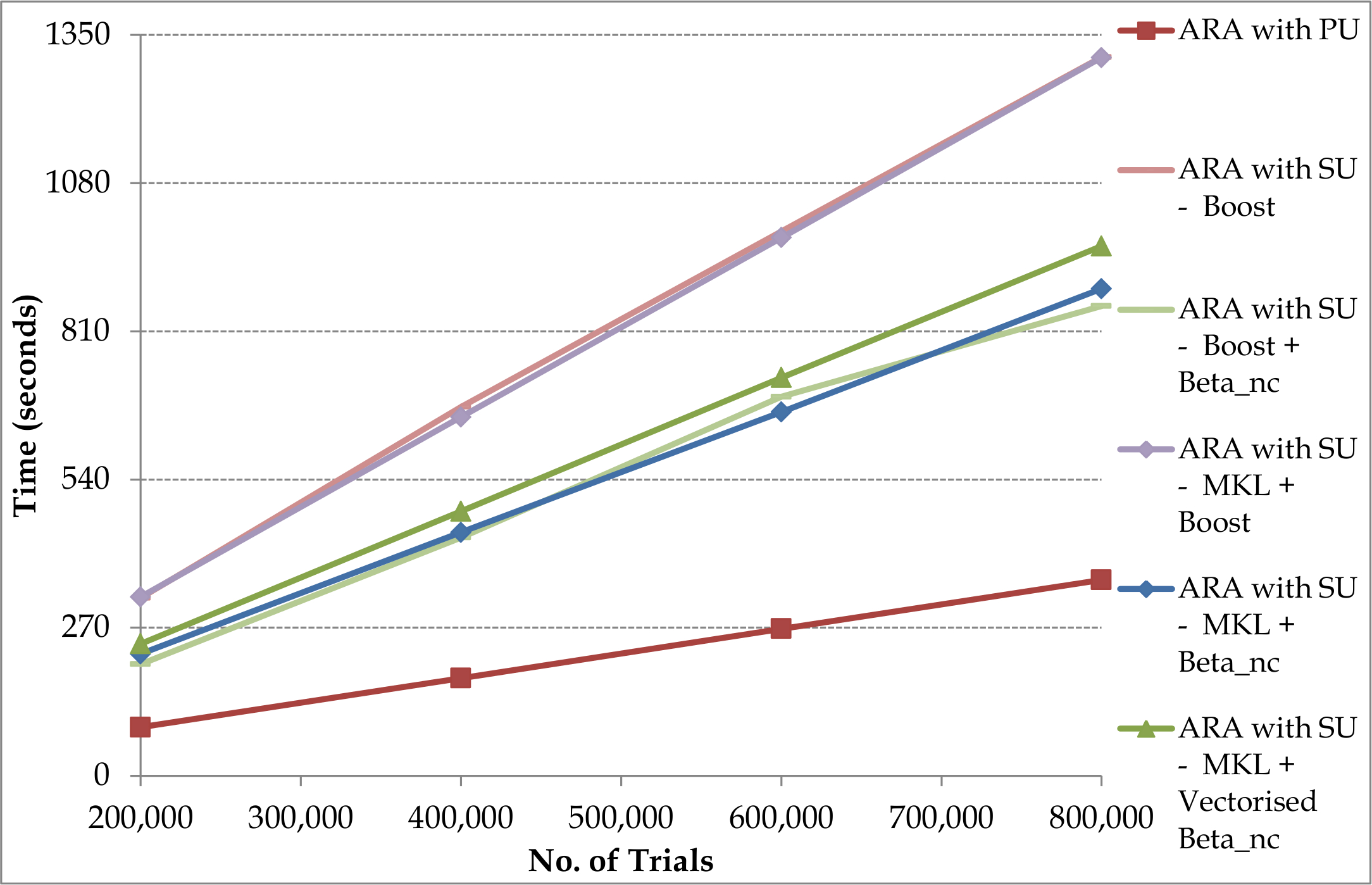}}
\caption{Sequential ARA with PU and SU for 200,000 to 800,000 trials using the Boost, MKL and Beta\_nc libraries on CPU}
\label{figure1}
\end{figure}

Figure \ref{fig:1a} and Figure \ref{fig:1b} show the results from the i7 and Xeon CPUs respectively, when the Boost, MKL and Beta\_nc libraries are employed to sequentially perform ARA. 
In the best case, the i7 is nearly 49\% faster than Xeon for ARA with PU and 95\% faster than Xeon for ARA with SU, resulting in the i7 being 1.32 times faster than Xeon. 

Surprisingly, the combination of MKL and the vectorised Beta\_nc libraries are outperformed by the combination using MKL and non-vectorised Beta\_nc library. 
The vectorised code performs well when the precision of the inverse Beta CDF is high. 
In all the experiments presented in this paper, the solution for the inverse Beta CDF converges when there is accuracy for up to six decimal places (or relative error is less than $10^{-6}$). 
However, if the precision of the solution is increased for greater accuracy, for example, twelve decimal places, then the vectorised library improves the performance significantly. 
When the precision is low only few iterations are required for the solution to converge; few iterations cannot reap the benefit of vectorisation, but introduces overheads which is a trade-off. 

All possible combinations of the libraries for the CDF and the Quantile of the CDF for the Normal distribution and the Quantile of the Beta distribution were employed. On both the i7 and Xeon, the combination of the MKL (for the Normal distribution functions) and Beta\_nc (for the Beta distribution function) libraries gave the best results. One advantage of using the MKL library is that it supports the Intel AVX instructions and performs well on Intel hardware.

\begin{figure} %[t][!tp]
\centering
	\subfloat[i7]{\label{fig:2a}\includegraphics[width=0.5\textwidth]{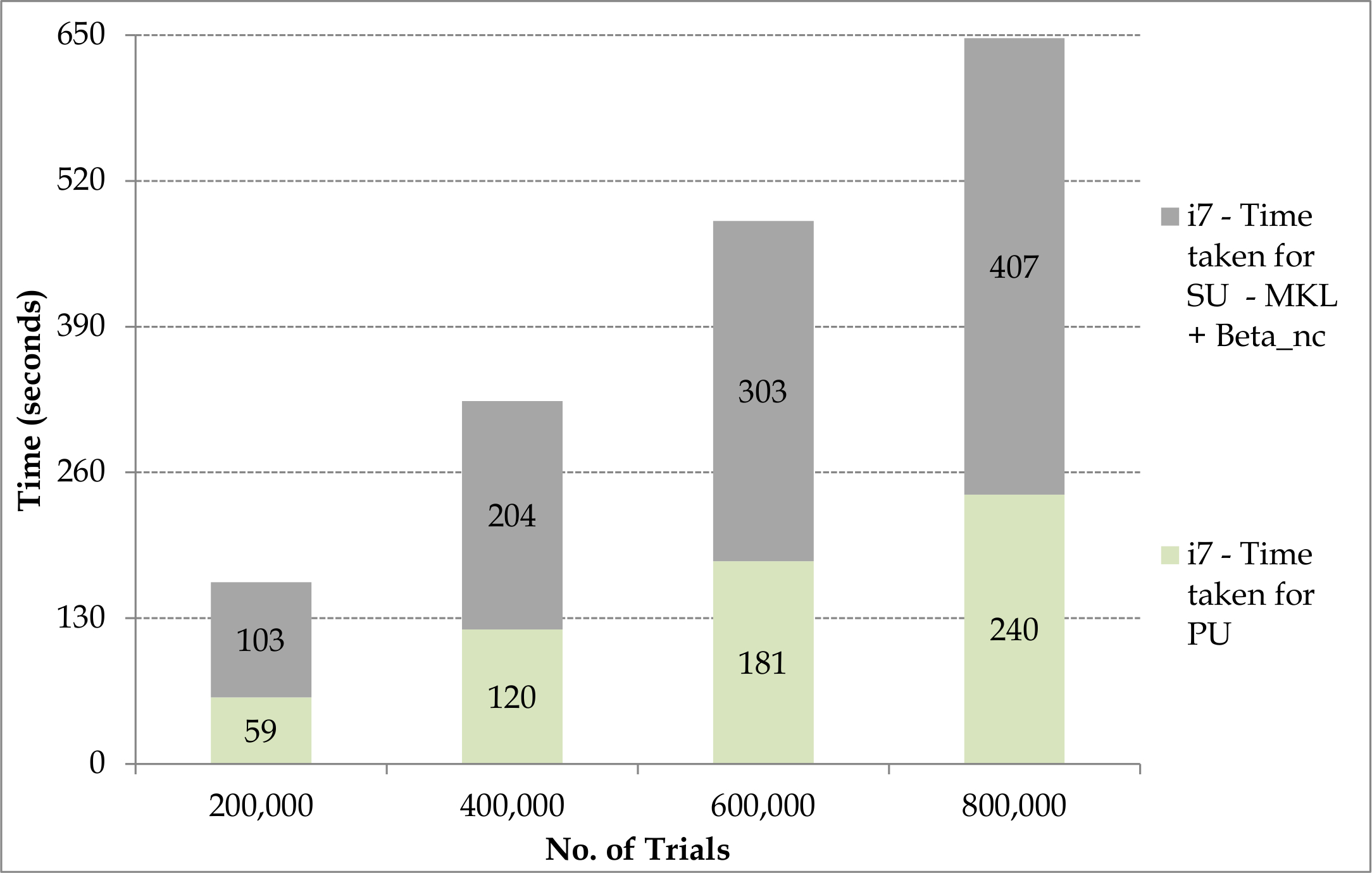}} \hspace{12pt}
	\subfloat[Xeon]{\label{fig:2b}\includegraphics[width=0.5\textwidth]{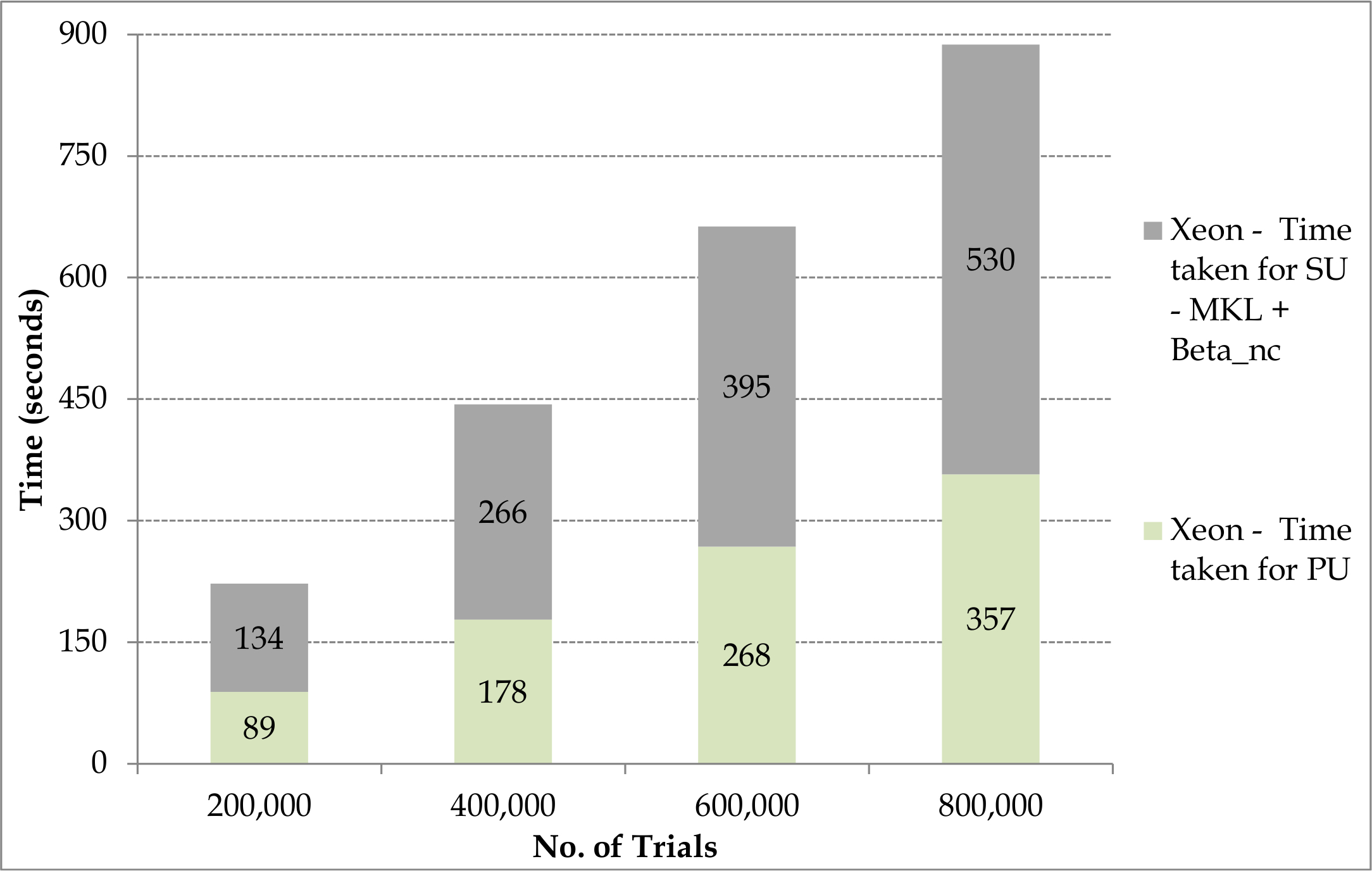}}
\caption{Time taken for sequential ARA with PU and SU using the Boost library on CPU}
\label{figure2}
\end{figure}

Figure \ref{figure2} shows the individual times taken to apply PU and SU on the i7 and Xeon CPUs when Boost library is used. 
The time for applying SU is nearly 2.5 times the time taken for ARA in each case of trials shown in the graph. 
The mathematical functions employed for applying SU are fast methods, although the inverse CDF of the beta distribution which takes majority of the time is an exception; over 95\% of the time taken for SU is required by the inverse CDF. 
The time for applying secondary uncertainty is nearly 1.5 - 1.7 times the time taken for ARA in each case of trials shown in the graph. 

The time taken for ARA with PU and SU with increasing number of trials should scale linearly. This is observed both in Figure \ref{figure1} and Figure \ref{figure2}. 
%===========================================================================

\subsection{Parallel Implementations}

The results obtained from parallel implementations on the multi-core CPUs are shown in Figure \ref{figure3} to Figure \ref{figure5}. The results when many core hardware accelerators are used in the native mode or individually are shown in Figure \ref{figure6} to Figure \ref{figure11}. Results obtained on hybrid platforms comprising multi-cores and many-cores are shown in Figure \ref{figure12} to Figure \ref{figure9}. 

%===========================================================================
%-------------------------------OPENMP-PARALLEL-----------------------------
%===========================================================================

\subsubsection{On CPU}
\begin{figure} %[t][!tp]
\centering
	\subfloat[i7]{\label{fig:3a}\includegraphics[width=0.5\textwidth]{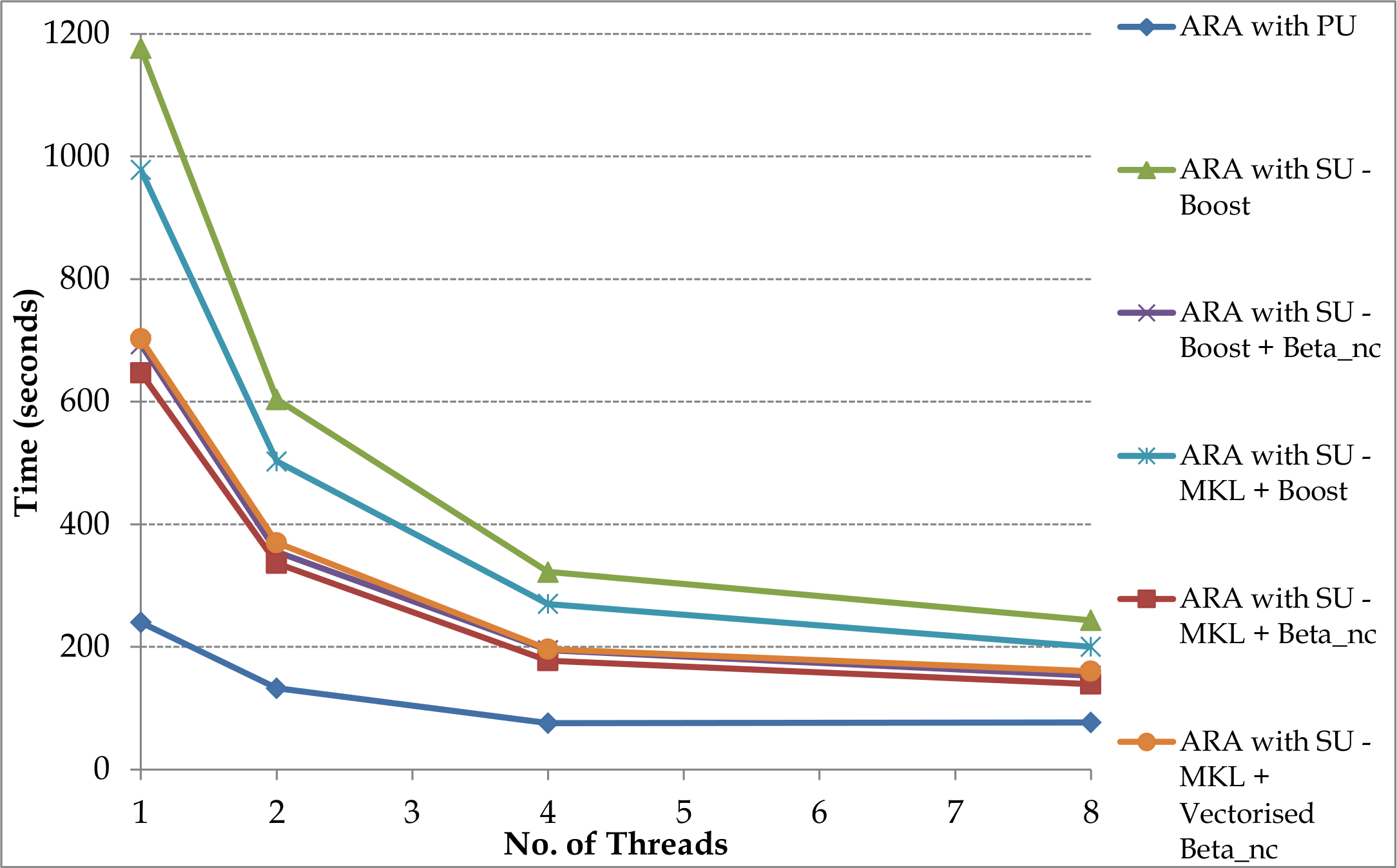}} \\
	\subfloat[Xeon]{\label{fig:3b}\includegraphics[width=0.5\textwidth]{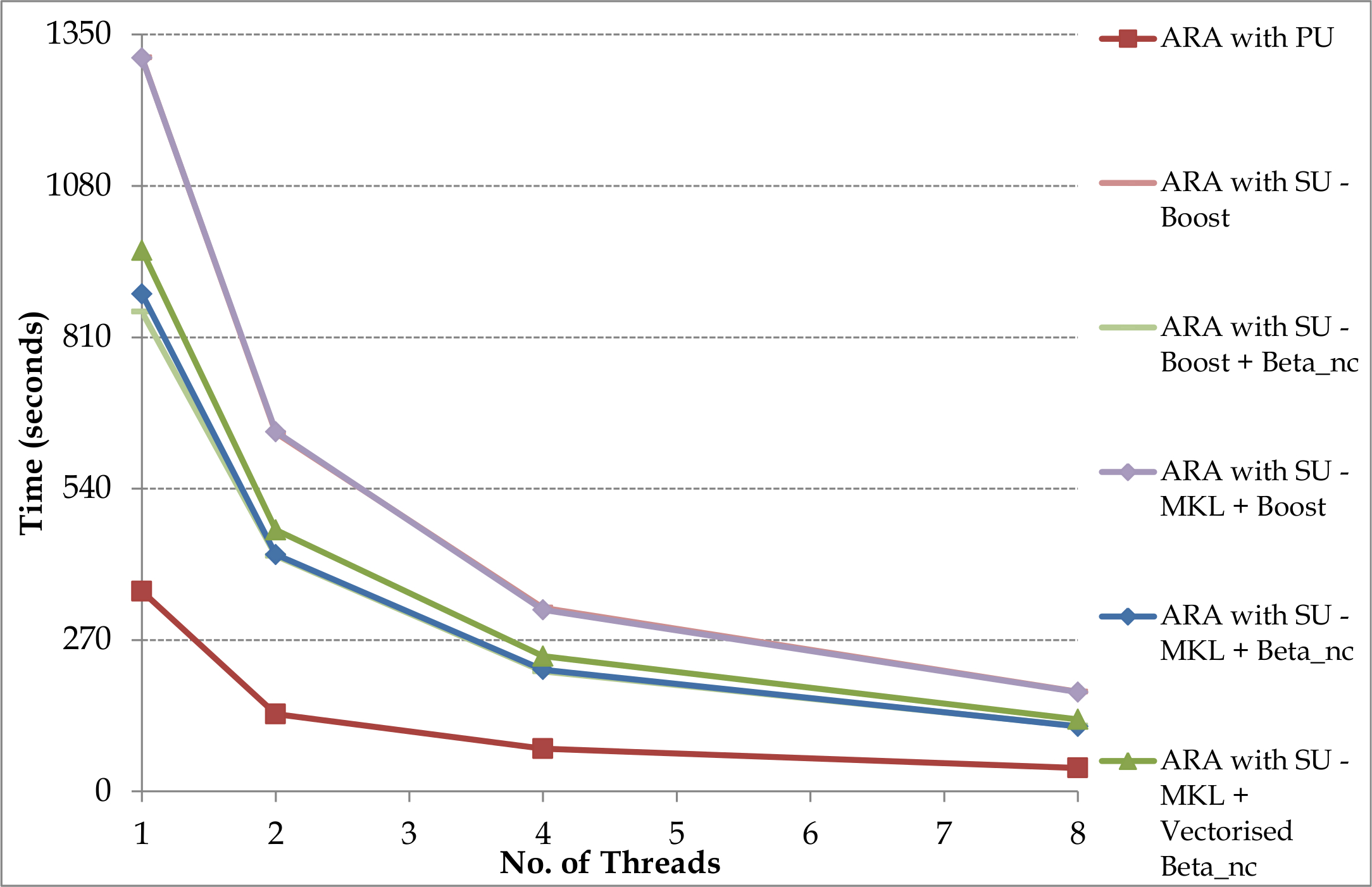}}
\caption{Time taken for ARA with PU and SU using one thread per computing core on multi-core CPUs for 800,000 trials}
\label{figure3}
\end{figure}

%Figure 3a and 3b
Figure \ref{figure3} shows the time taken for performing ARA with PU and SU when the number of threads are varied from 1 to 8 on the i7 (Figure \ref{fig:3a}) and the Xeon (Figure \ref{fig:3b}) platforms using the Boost, MKL and Beta\_nc libraries for 800,000 trials. 
A single thread is run on each core of the CPU and the number of cores are varied from 1 to 8. 
Each threads performs ARA with PU and SU for a single trial. 
Multiple threads are used by employing OpenMP directive \texttt{\small \#pragma omp parallel} in the C++ source. 
With respect to the overall time on the i7 and Xeon CPUs, the MKL and Beta\_nc combination performs the best requiring around 113 seconds.
On both platforms, the performance of the standalone Boost library is poorer than the combination of libraries. 
The vectorised library does not perform better than the unvectorised library. 
This is so, since the benefits of vectorisation are evident only when high precision is required from the solution of the inverse Beta CDF.
While there are only few iterations required for converging to six decimal place precision, as required in this research, the benefits of vectorisation are balanced off by the overheads in the vectorised library.

In the case of the best performer, the MKL and the Beta\_nc library, a speed up of 4.6x is obtained on i7 and 7.6x is obtained on Xeon. The performance on the Xeon is better due to the larger number of cores compared to the i7 and the bandwidth to memory on the Xeon which is 2.4 times that on the i7. The majority of the time taken to compute primary uncertainty is for performing random access reads into the data structure representing the XELT (this is further discussed in Section \ref{experimentalresults:discussion}; on the i7 and Xeon, the time taken for such memory operations is nearly five times the time taken for the remaining computations while applying financial terms in PU, also shown later in Figure \ref{figure13}). The majority of the time for applying secondary uncertainty in ARA is consumed in the inverse cumulative distribution function of the beta distribution (nearly 95\% of the time taken for SU). 

\begin{figure} %[t][!tp]
\centering
	\subfloat[i7]{\label{fig:4a}\includegraphics[width=0.5\textwidth]{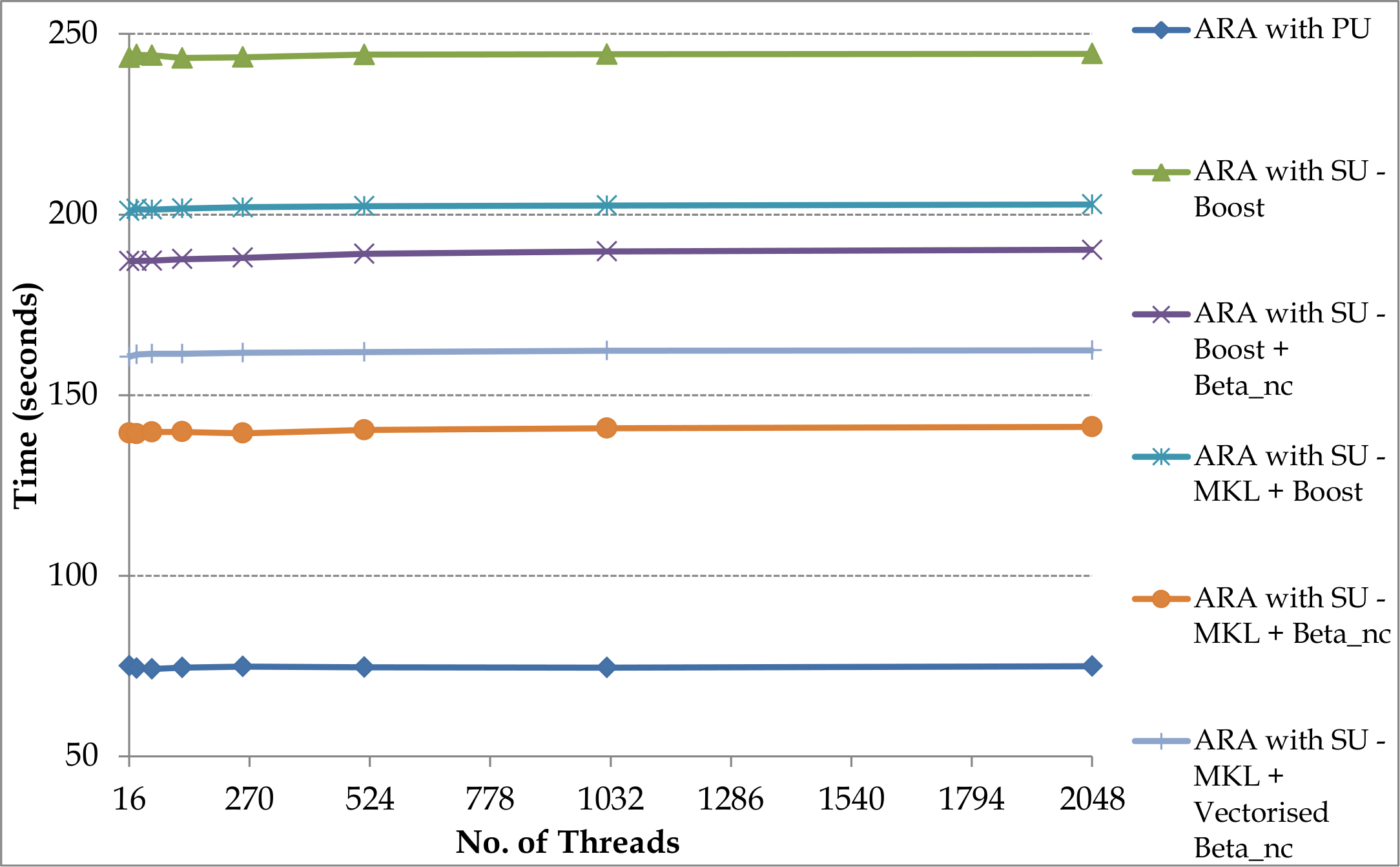}} \\
	\subfloat[Xeon]{\label{fig:4b}\includegraphics[width=0.5\textwidth]{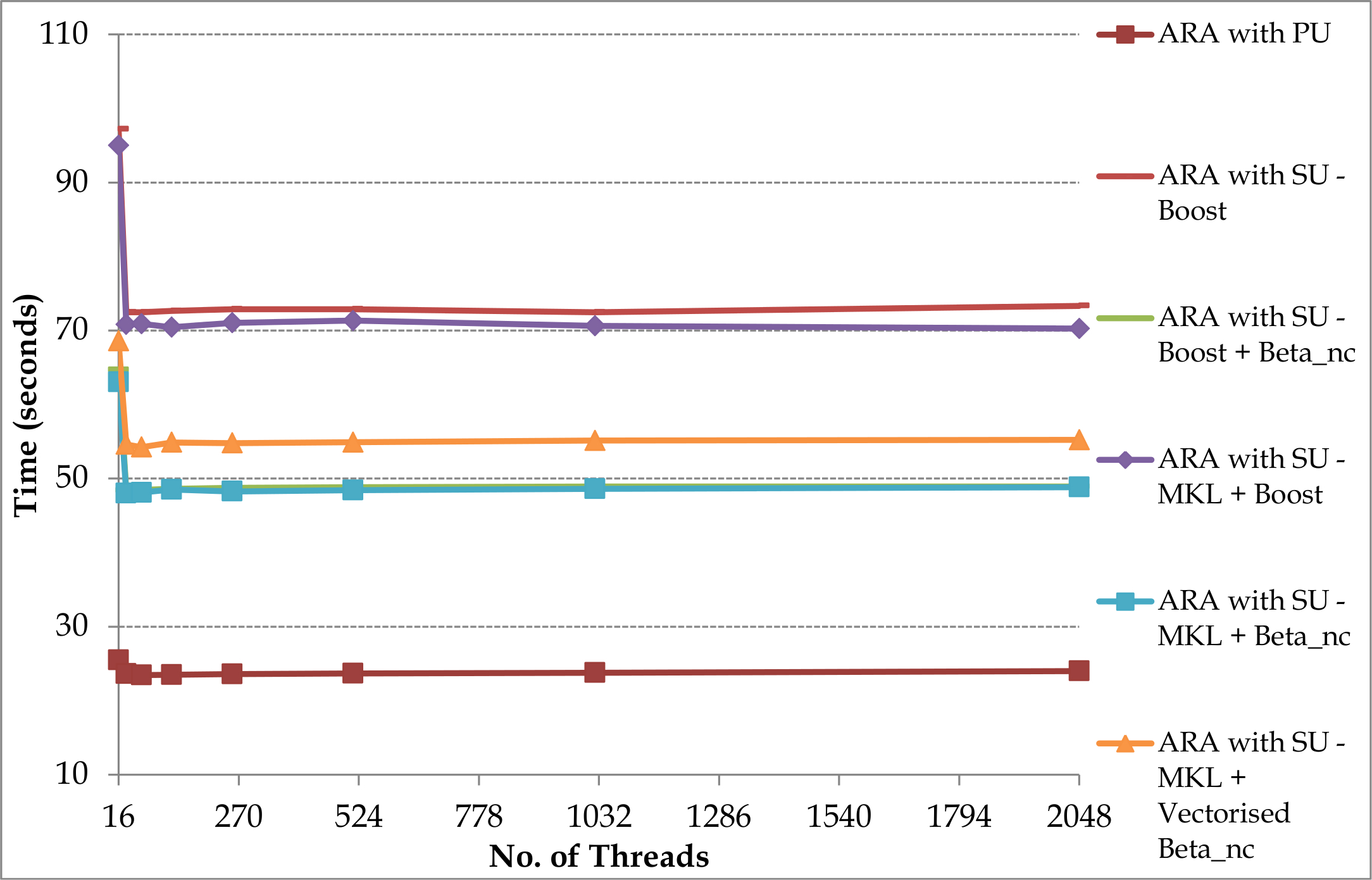}}
\caption{Time taken for ARA with PU and SU using multiple thread per computing core on multi-core CPUs for 800,000 trials}
\label{figure4}
\end{figure}

%Figure 4a and 4b
Figure \ref{figure4} shows the graphs plotted for the time taken for performing ARA accounting both for PU and SU when the number of threads are varied from 16 to 2048 on the i7 (Figure \ref{fig:4a}) and Xeon (Figure \ref{fig:4b}) platforms using the Boost, MKL and Beta\_nc libraries for 800,000 trials. 
Multiple threads are run on each core of the CPU with the exception of the Xeon until 32 threads are used. 
For example, when 16 threads are employed on the i7 and Xeon two threads and one thread run on each core respectively, and when 2048 threads are employed on the i7 and Xeon 256 threads and 128 threads run on each core respectively. 
In practice, the cores are not generally over subscribed, but is presented in the experiments to show the effect of over subscription.
For example, super linear speed ups are observed due to hyperthreading. 
On the i7 a small drop is observed first in the absolute time followed by a slight but gradual increase when many threads are executed on each core. 
However, on the Xeon, before there is a similar observation, there is a steep drop in the time moving from 16 to 32 threads. 
This is because the Xeon has 16 cores and the timings start to stagnate only after all the 16 cores are used. 
On the i7 the best result is 139 seconds compared to 48 seconds on the Xeon; the better performance by a factor of 3 is attributed to twice the number of cores and 2.6 times the bandwidth to memory compared to the i7.

\begin{figure} %[t][!tp]
\centering
	\subfloat[i7]{\label{fig:5a}\includegraphics[width=0.5\textwidth]{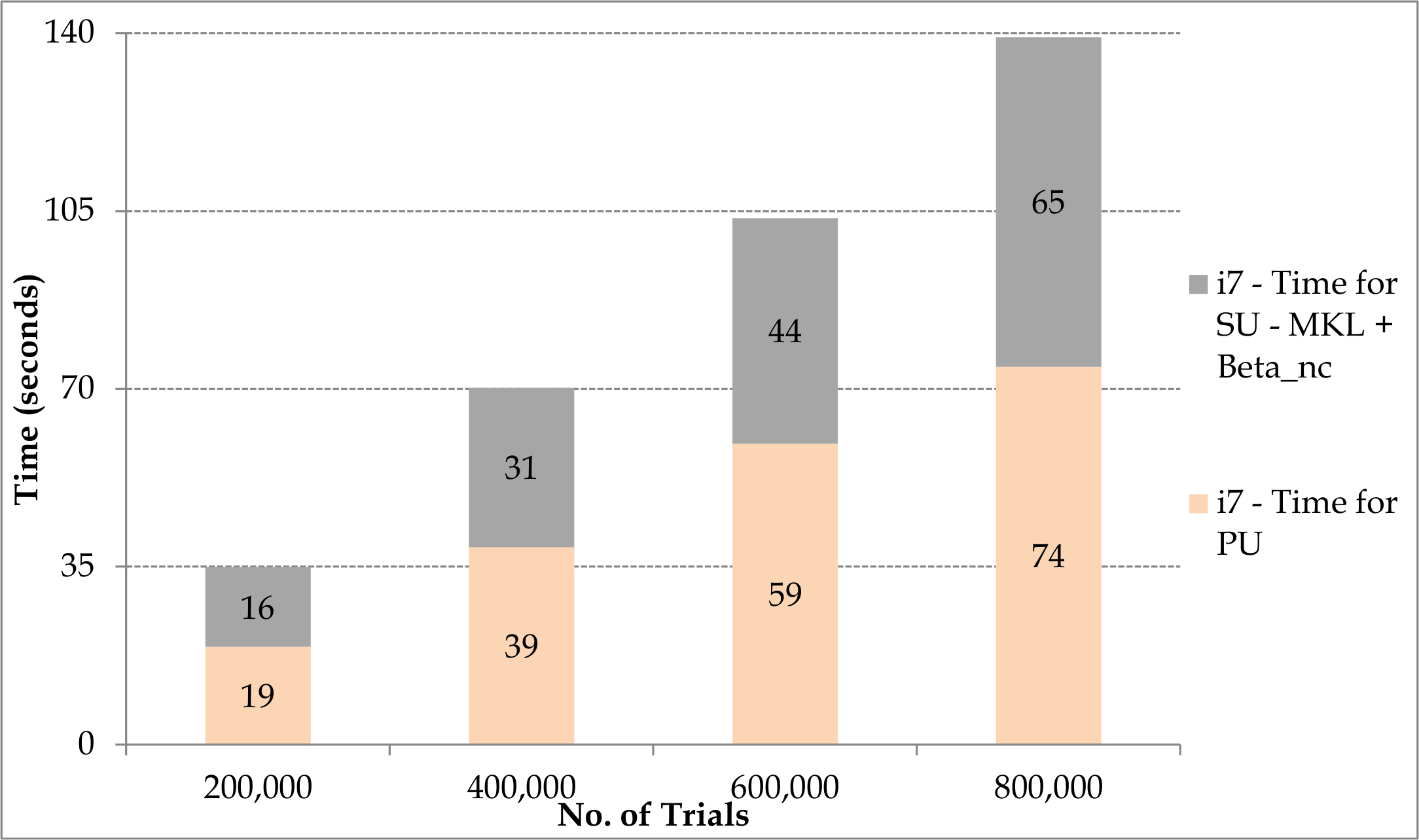}} \\
	\subfloat[Xeon]{\label{fig:5b}\includegraphics[width=0.5\textwidth]{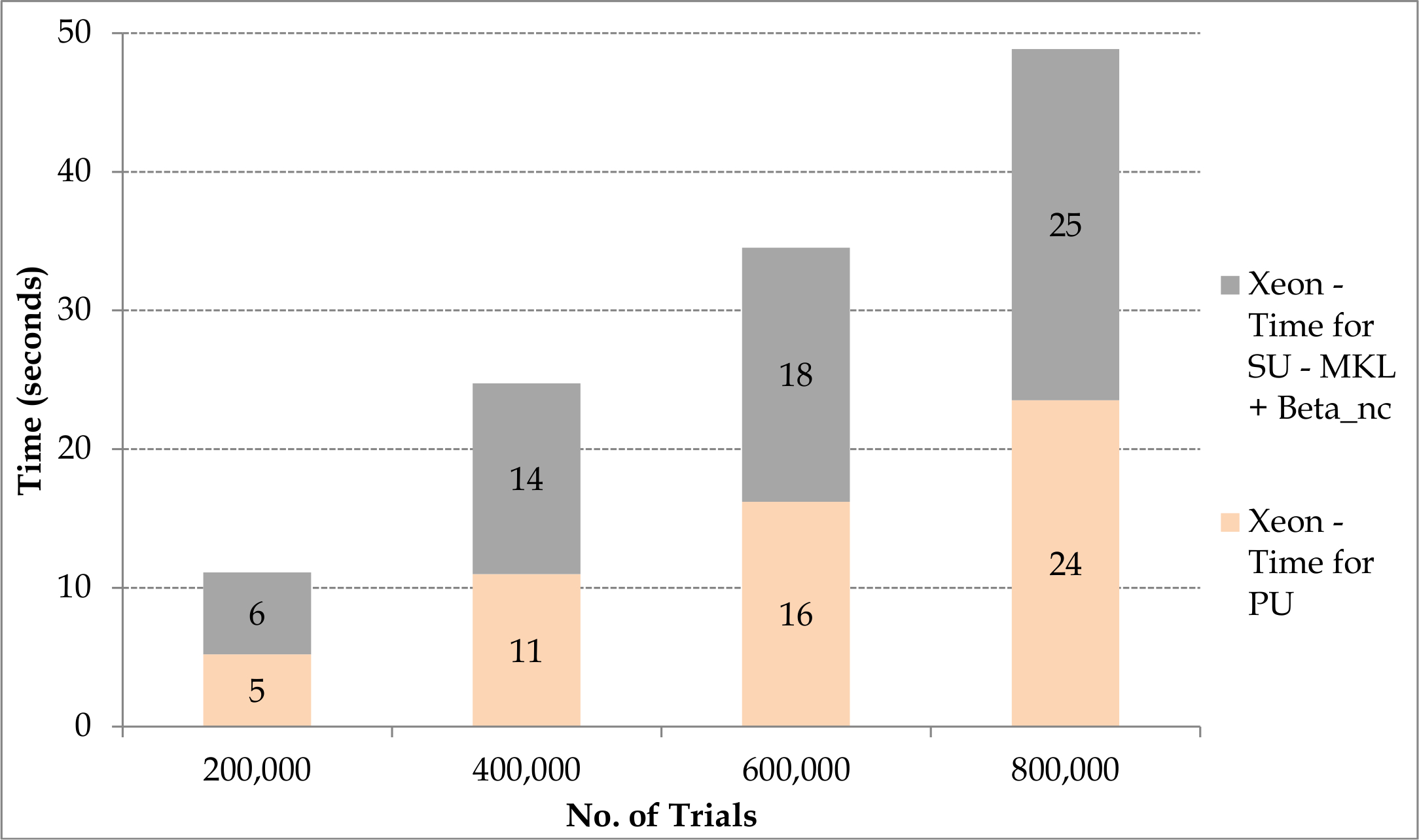}}
\caption{Time taken for ARA with PU and SU using MKL and Beta\_nc libraries using 32 threads on multi-core CPUs}
\label{figure5}
\end{figure}

%Figure 5a and 5b
Figure \ref{figure5} shows the graphs plotted for the time taken for ARA with PU and SU on the i7 and Xeon platforms for 32 threads (best performance is obtained for 32 threads on both platforms). The time taken for applying SU increases with the number of trials in each case of trials shown in the graph. The time taken for applying SU on the i7 is only $\frac{1}{6}^{th}$ the time taken in the sequential implementation on the i7. The time taken for ARA with SU on the Xeon is only $\frac{1}{21}^{th}$ the time taken in the sequential implementation. The Xeon is nearly three times faster than the i7 for applying PU and upto 2.6 times faster for applying SU. Overall a speed up of approximately 5x and 18x is obtained on the i7 and Xeon over their respective sequential implementations.

%===========================================================================

\subsubsection{On Hardware Accelerators only}

%===========================================================================
%------------------------------------GPU------------------------------------
%===========================================================================

\begin{figure} %[t][!tp]
\centering
	\subfloat[For 800,000 trials using varying threads per block]{\label{fig:6a}\includegraphics[width=0.5\textwidth]{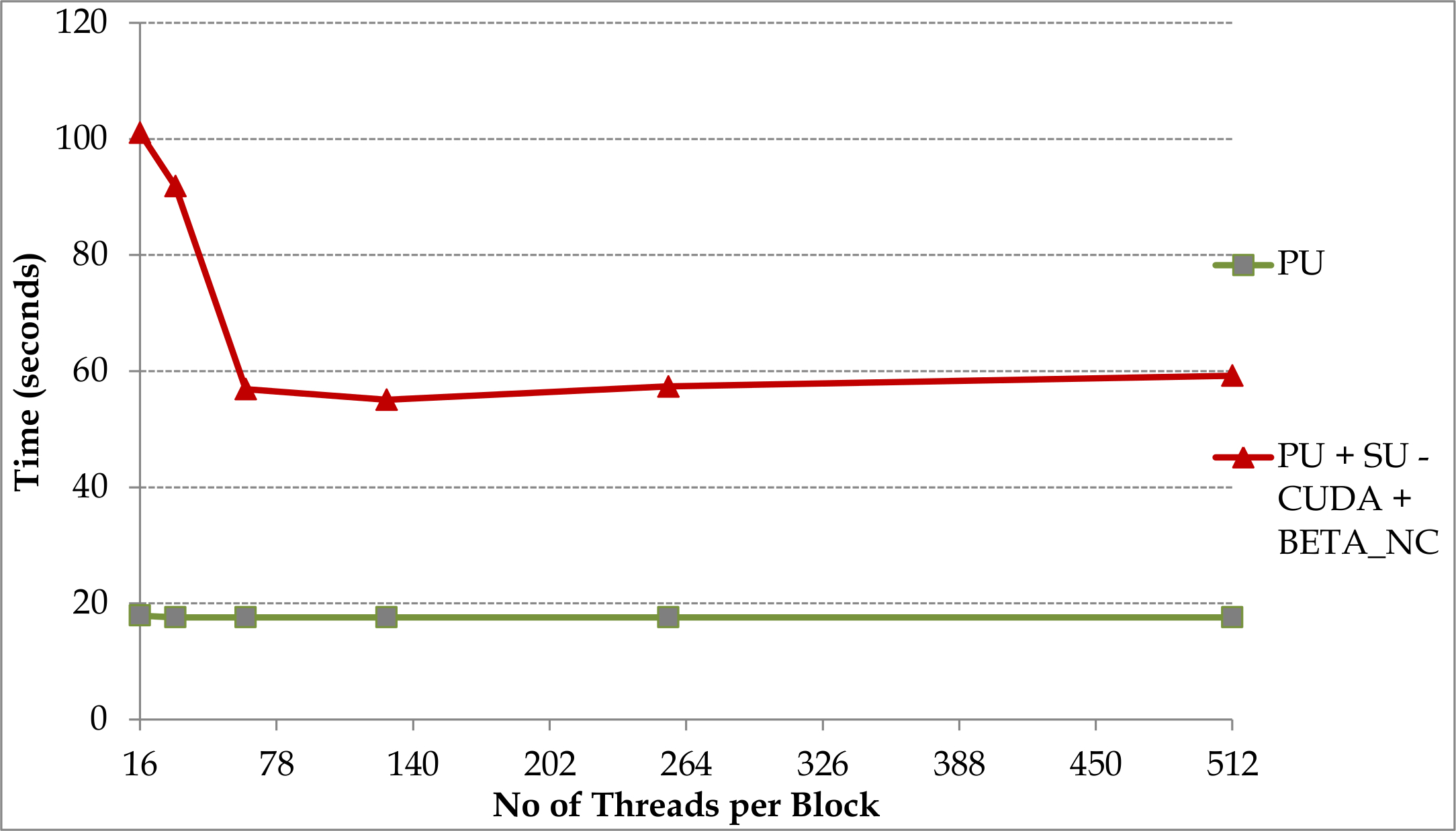}} \\
	\subfloat[Using 128 threads per block for varying number of trials]{\label{fig:6b}\includegraphics[width=0.5\textwidth]{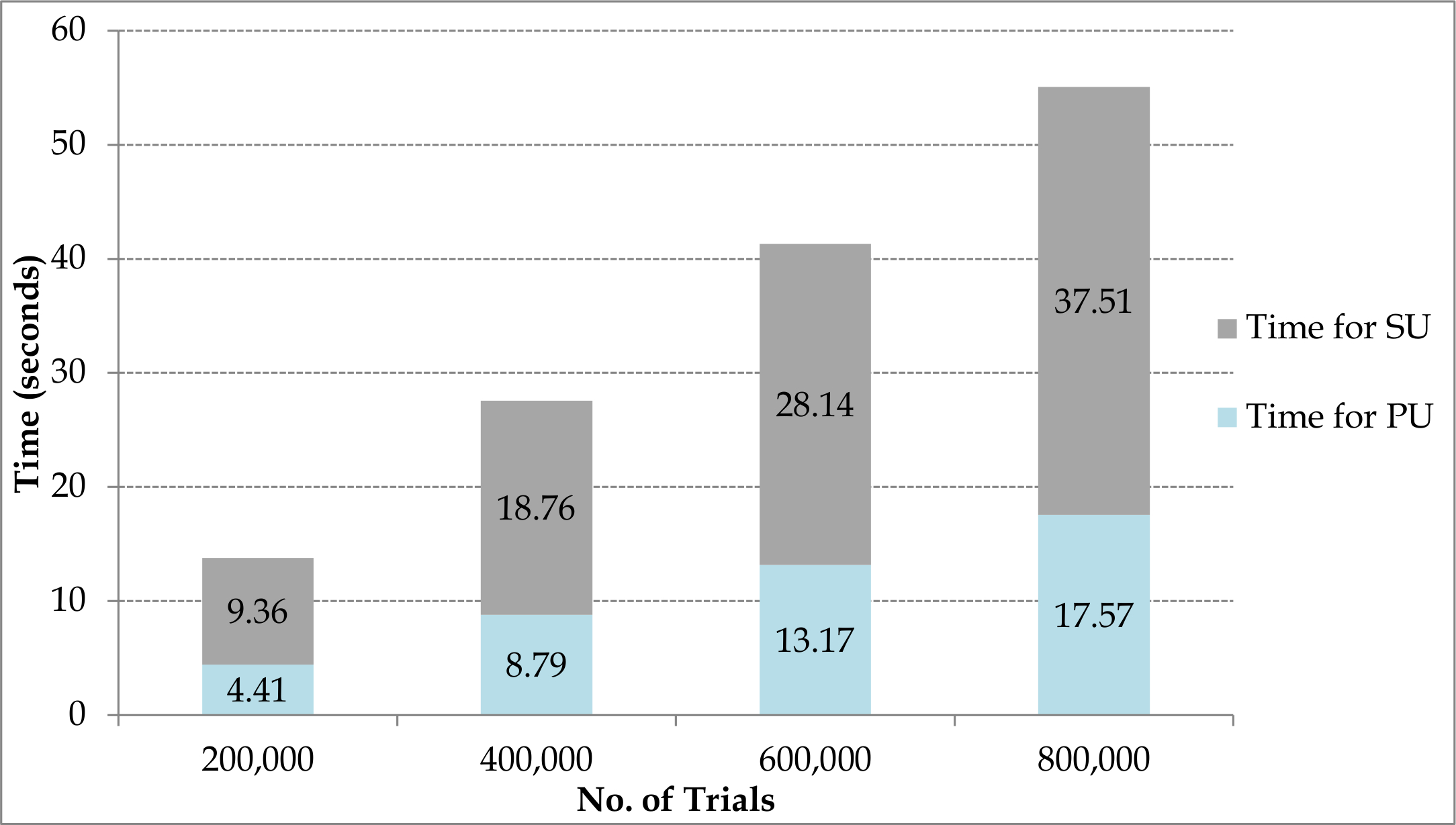}}
\caption{Time taken for ARA on the GPU}
\label{figure6}
\end{figure}

Figure \ref{fig:6a} and Figure \ref{fig:6b} show the graphs plotted for the time taken for performing ARA considering both PU and SU for 800,000 trials on the many-core GPU using the CUDA Math and Beta\_nc libraries. The Boost library is not available for GPUs. 

In Figure \ref{fig:6a}, the analysis is performed for 800,000 trials on the GPU by varying the number of threads per block from 16 to 512 threads. 
An improvement in the performance is seen as the number of threads per block increase from 16 to 128 since the latency for accessing the global memory drops. 
Beyond 128 threads, the performance starts to drop, since the shared and constant memory available to each thread decreases. 
The lowest time taken is 55 seconds when 128 threads per block are used, which is nearly 12 times faster than the best sequential performance on i7 and 15 times faster than the best sequential performance on Xeon. 
The overall time for the analysis on the GPU is comparable to the performance of the multi-threaded implementation of 2048 threads on the Xeon.

Figure \ref{fig:6b} shows the time taken for performing ARA incorporating PU and SU using 128 threads per block on the GPU. Both times scale linearly as expected. The time taken for applying SU is over twice the time taken for performing ARA.

%===========================================================================

%------------------------------------PHI------------------------------------

\begin{figure} %[t][!tp]
\centering
	\subfloat[Single thread per core]{\label{fig:10a}\includegraphics[width=0.5\textwidth]{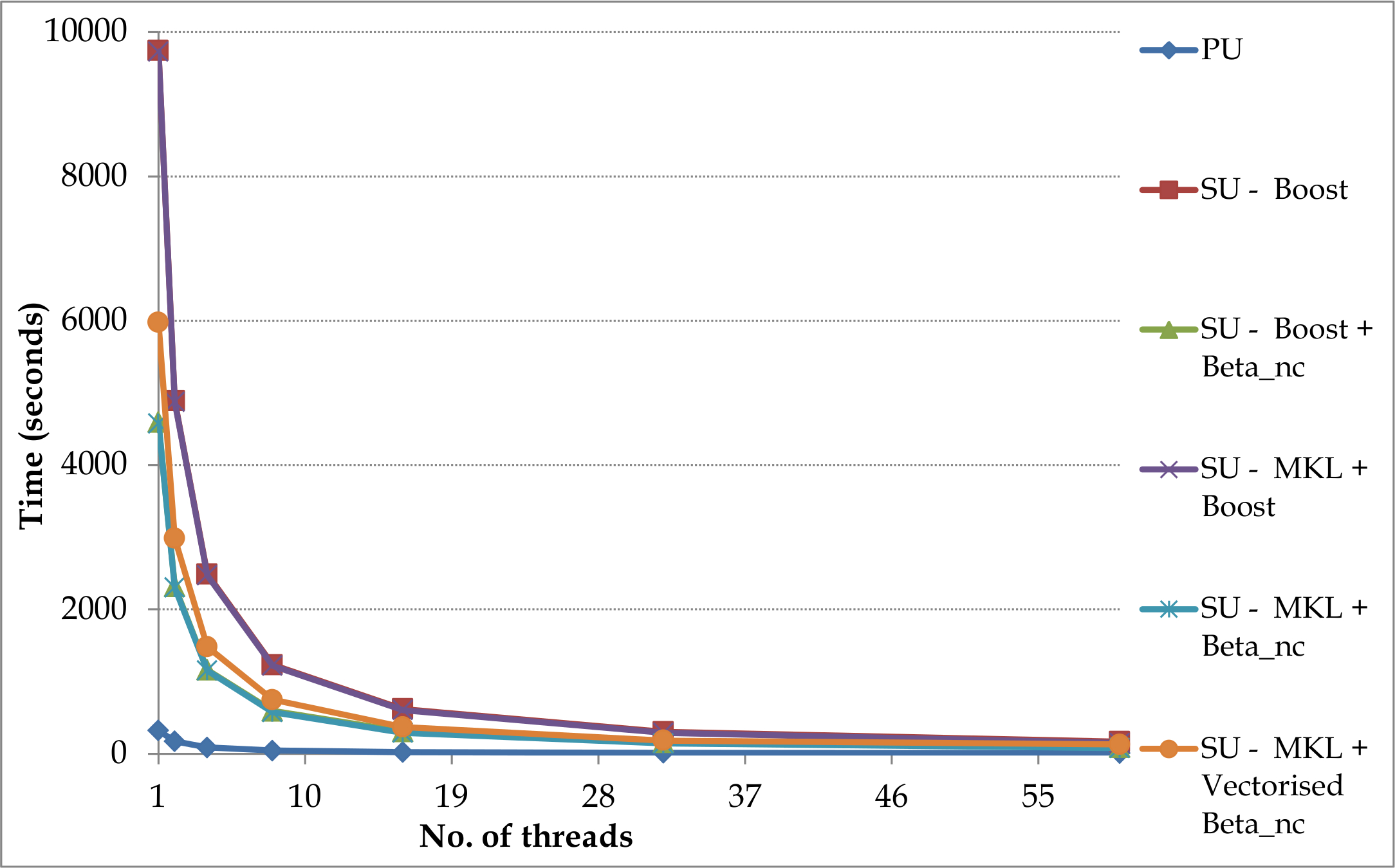}} \\
	\subfloat[Multiple threads per core]{\label{fig:10b}\includegraphics[width=0.5\textwidth]{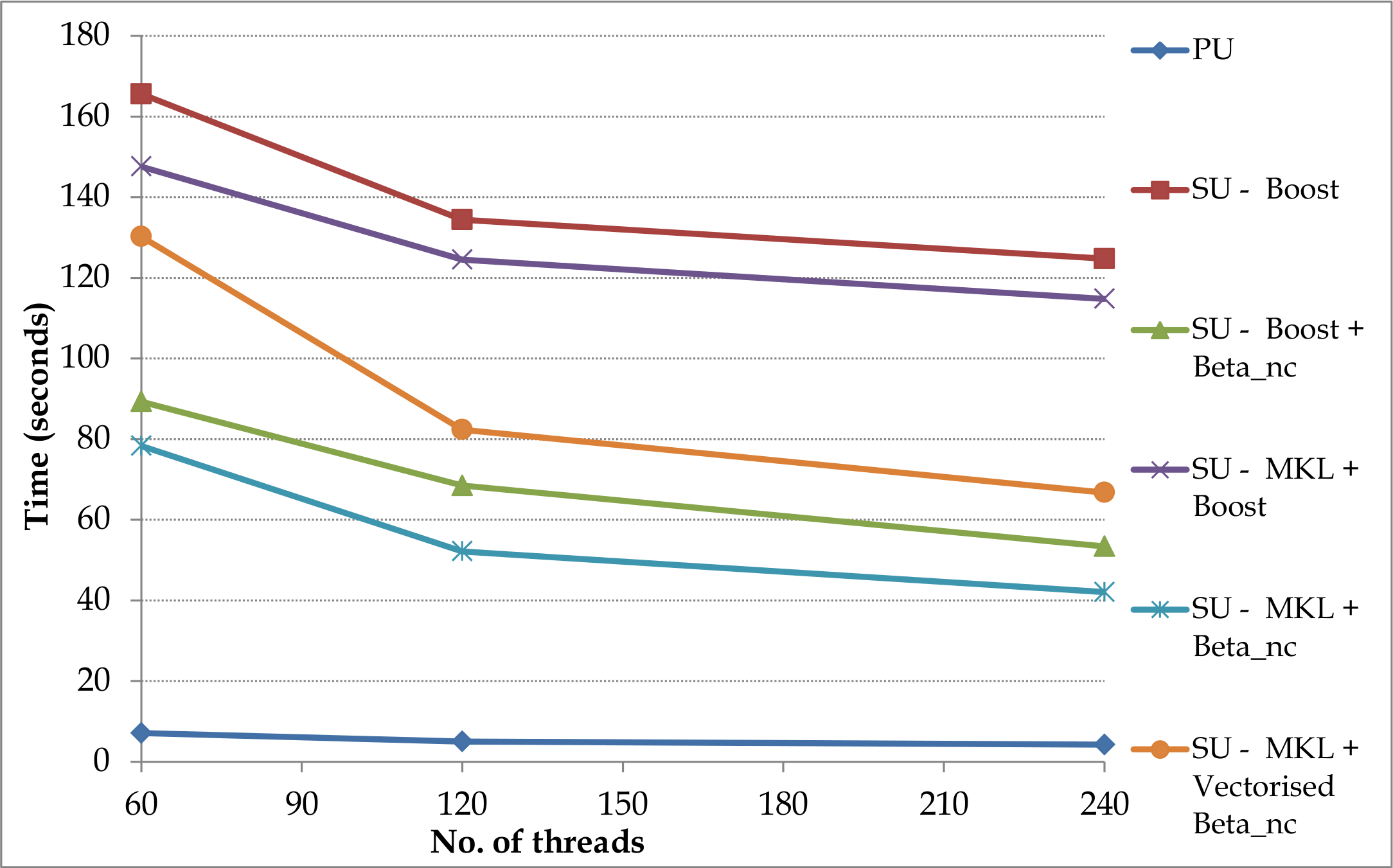}}
\caption{Time taken for ARA with PU and SU using Intel Phi for 800,000 trials}
\label{figure10}
\end{figure}

\begin{figure}
	\centering
	\includegraphics[width = 0.5\textwidth]{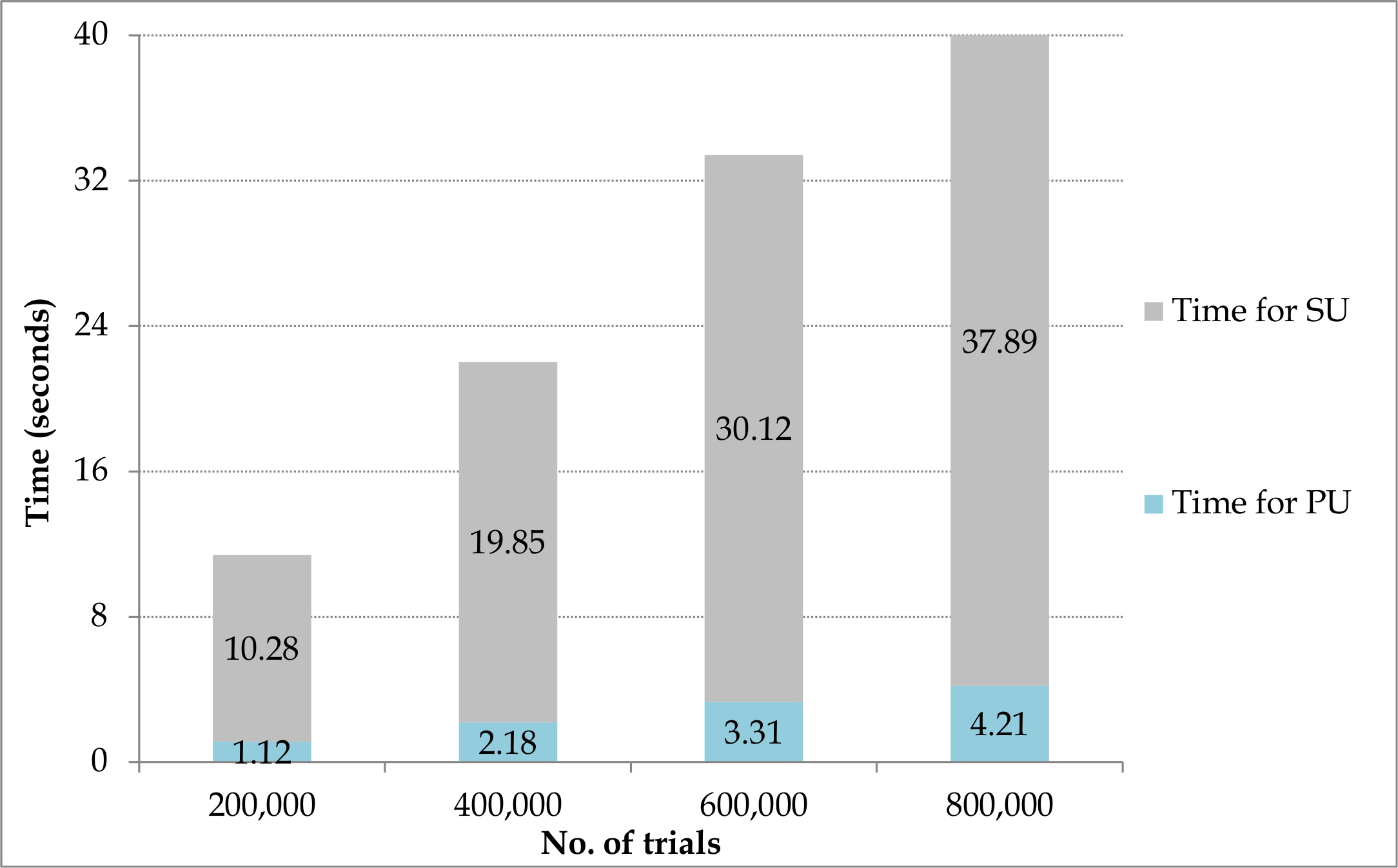}
	\caption{Times taken for applying PU and SU in ARA on the Phi for varying number of trials}
	\label{figure11}
\end{figure}

Figure \ref{figure10} and Figure \ref{figure11} show the experimental results obtained when the Intel Phi alone is employed for performing ARA with PU and SU using Boost, MKL and Beta\_nc libraries. 
Figure \ref{fig:10a} shows the performance of ARA when one thread is executed on each core of the Phi (the number of threads is varied from 1 to 60). 
As expected there is an increase in speed when the number of threads is increased. 
The fastest time obtained is 78.32 seconds on 60 threads using MKL and Beta\_nc. 
The combination of Boost and Beta\_nc also perform comparably. 
ARA with PU consists mostly of memory related operations and does not experience any significant speedup. 
For example, 12 seconds are required on 30 threads as against 7 seconds on 60 threads. 
Over 90\% of the time is required for applying secondary uncertainty when 60 threads are used. 
Vectorisation does not yield any potential benefit over non-vectorised functions in applying secondary uncertainty due to overheads; vectorised function is nearly 38\% slower than the fastest non-vectorised function on 60 threads.  

Figure \ref{fig:10b} shows the performance of ARA when PU and SU are applied using multiple threads per core on the Phi (number of threads is varied from 60 to 240). 
Speedup is observed although there is limited efficiency. 
Again best performance is noted when MKL and Beta\_nc libraries are employed. 
The analysis is completed in 52.16 seconds and 42.1 seconds when two threads per core and three threads per core are used respectively.

Figure \ref{figure11} shows that there is a linear increase in the time taken for applying PU and SU on the Phi processor for varying number of trials. The time taken for SU is over 8 times the time taken for PU; the for SU is just under 38 seconds both for the Phi and the GPU. The real difference is observed for PU in that the Phi is four times faster than the GPU for performing memory computations. 

%===========================================================================

\subsubsection{On Hybrid platforms}

%-------------------------------------GPU + i7-------------------------------------

\begin{figure}
	\centering
	\includegraphics[width = 0.5\textwidth]{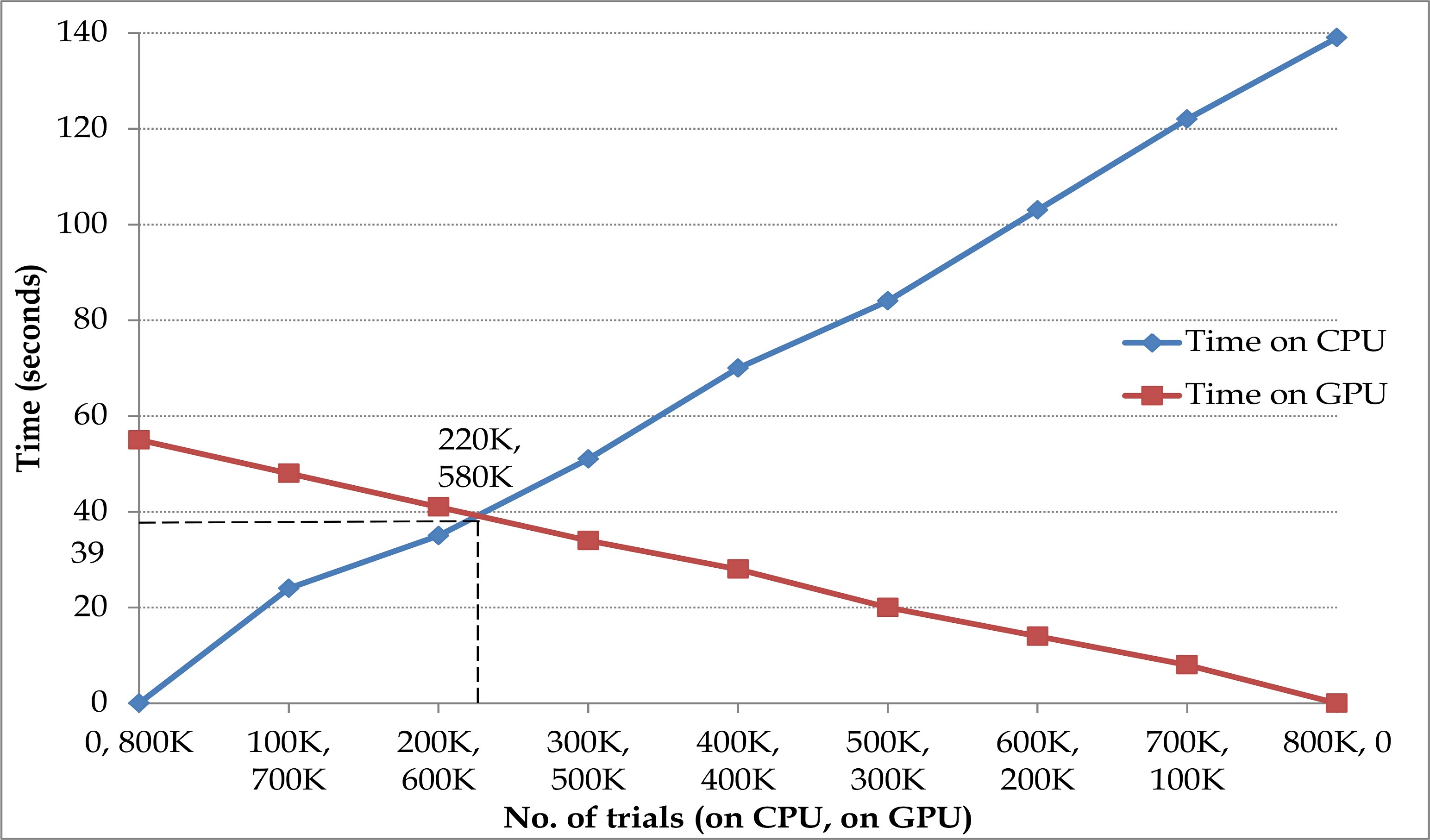}
	\caption{Times taken on the i7 CPU and Tesla C2075 GPU for applying PU and SU in ARA for varying trials}
	\label{figure12}
\end{figure}

Figure \ref{figure12} shows the time taken for performing ARA on the i7 CPU and the Tesla C2075 GPU used as a hybrid platform. 
The workload of 800,000 trials is split between the CPU and the GPU, with the intention that the CPU does not remain idle, in contrast to when the GPU alone is employed. 
The CPU makes use of the multi-cores of the processor and 16 threads using OpenMP to execute the workload; the MKL and Beta\_nc libraries are employed for applying secondary uncertainty. 
The workload on the GPU makes use of the many cores and 128 threads per core; the combination of the CUDA Math and Beta\_nc libraries are made use of to apply secondary uncertainty. 
The x-axis of the graph shows the varying workload of trials on the CPU and the GPU and the corresponding time taken for the workload on each platform. 
The best case scenario is when the CPU and the GPU take similar times for execution, which is the point of inflection on the graph. 
The CPU tends to be idle before the plots converge and the GPU is not utilised beyond the point of convergence. 
The optimal workload is when the CPU performs the analysis for 220,000 trials and the GPU performs the analysis on the remaining 580,000 trials. The analysis on the hybrid platform is 25\% faster (41 seconds) than the GPU (55 seconds). 

%===========================================================================

%------------------------------------XEON + PHI------------------------------------

\begin{figure} %[t][!tp]
\centering
	\subfloat[Single thread per core]{\label{fig:8a}\includegraphics[width=0.5\textwidth]{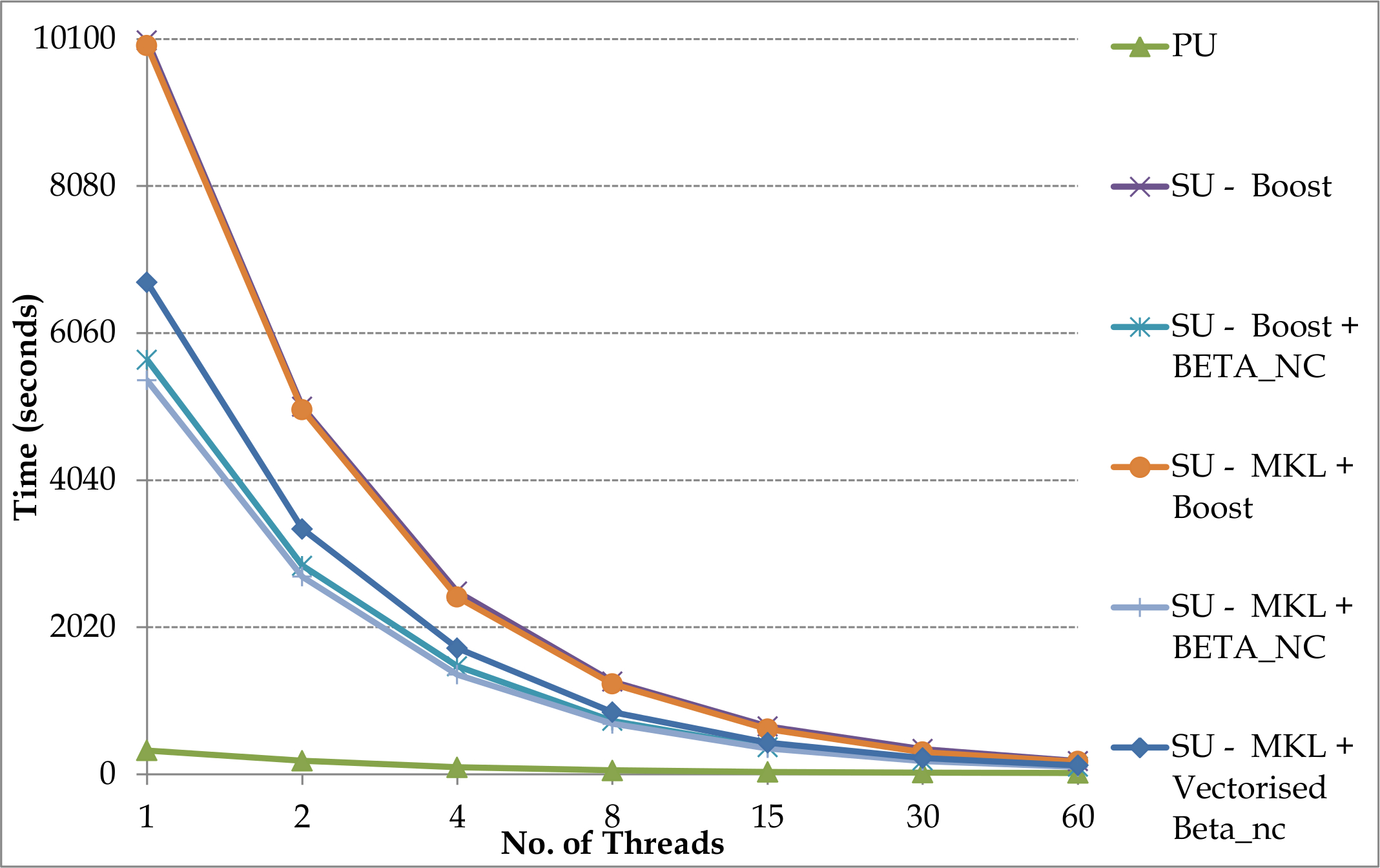}} \\
	\subfloat[Multiple thread per core]{\label{fig:8b}\includegraphics[width=0.5\textwidth]{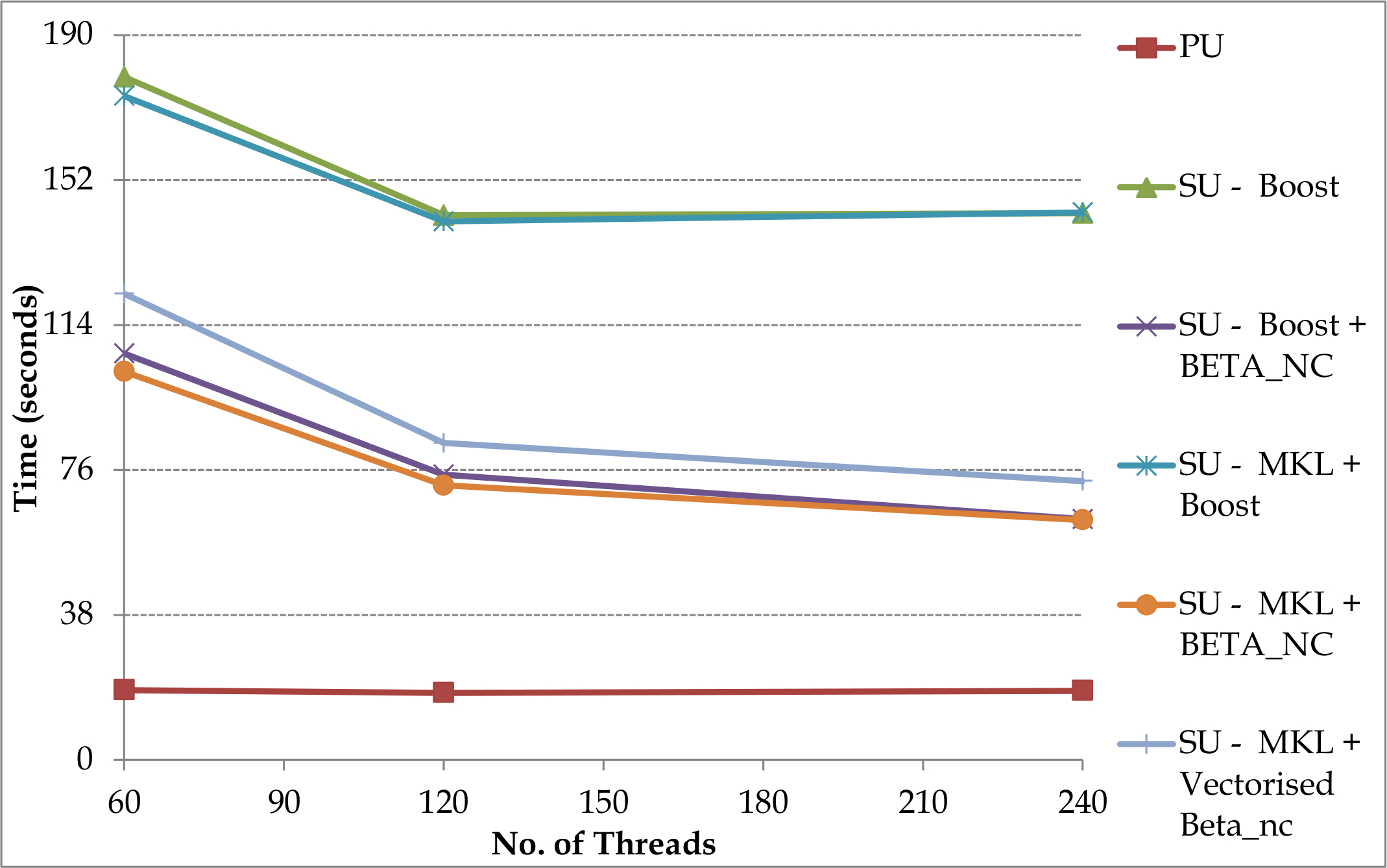}}
\caption{Time taken for ARA with PU and SU using Intel Xeon and Intel Phi Coprocessor}
\label{figure8}
\end{figure}

Figure \ref{figure8} shows the experimental results obtained when the Intel Xeon and Phi Coprocessor are employed for performing ARA accounting for both PU and SU using the Boost, MKL and Beta\_nc libraries. 
Figure \ref{fig:8a} is the graph showing the performance of ARA when the number of threads are varied from 1 to 60 (i.e., execution of single thread per core). 
As expected an increase in speed is noted with increasing number of threads. 
The fastest time obtained for ARA with SU is less than 102 seconds on 60 threads using the MKL and Beta\_nc libraries; the combination of the Boost and Beta\_nc libraries is also comparable.
ARA accounting only for PU does not enjoy the same degree of speedup, since it consists mostly of memory operations; 18 seconds are required on 60 threads and 24 seconds on 30 threads. 
Nearly 82\% percent more time is required for SU using 60 threads. Despite the increased potential for vectorization on the Phi coprocessor, the overhead in the vectorized functions causes it to be nearly 20\% slower than the non-vectorised library. 

Figure \ref{fig:8b} is the graph showing the performance of ARA incorporating PU and SU when the number of threads are varied from 60 to 240 (i.e., execution of multiple threads per core) on the Phi coprocessor. The speedup steadily decreases when multiple threads are executed on each core. For example, when two threads are executing per core the best time is 72 seconds compared to 62.88 seconds when three threads are executing per core. 

\begin{figure}
	\centering
	\includegraphics[width = 0.5\textwidth]{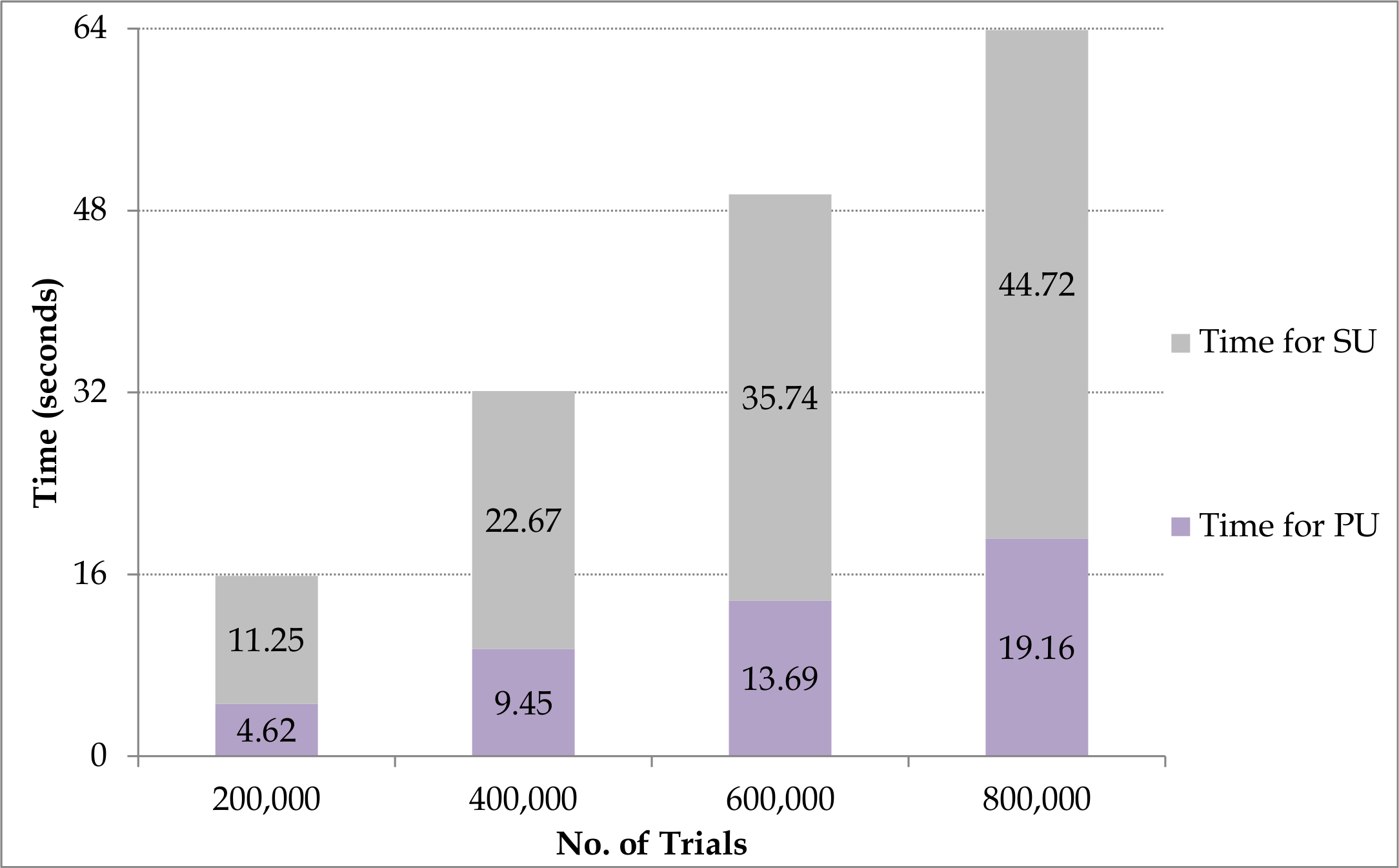}
	\caption{Times taken for applying PU and SU in ARA for varying number of trials on Xeon and Phi}
	\label{figure9}
\end{figure}

Figure \ref{figure9} shows the time taken for applying PU and SU on the Xeon and Phi. Both times scale linearly as expected. The time taken for applying SU is over twice the time taken for performing ARA with PU. The best library combination of MKL and Beta\_nc takes approximately 3.5 times longer for SU than for applying primary uncertainty. 

%===========================================================================

\subsection{Discussion}
\label{experimentalresults:discussion}

\begin{figure*}
	\centering
	\includegraphics[width = \textwidth]{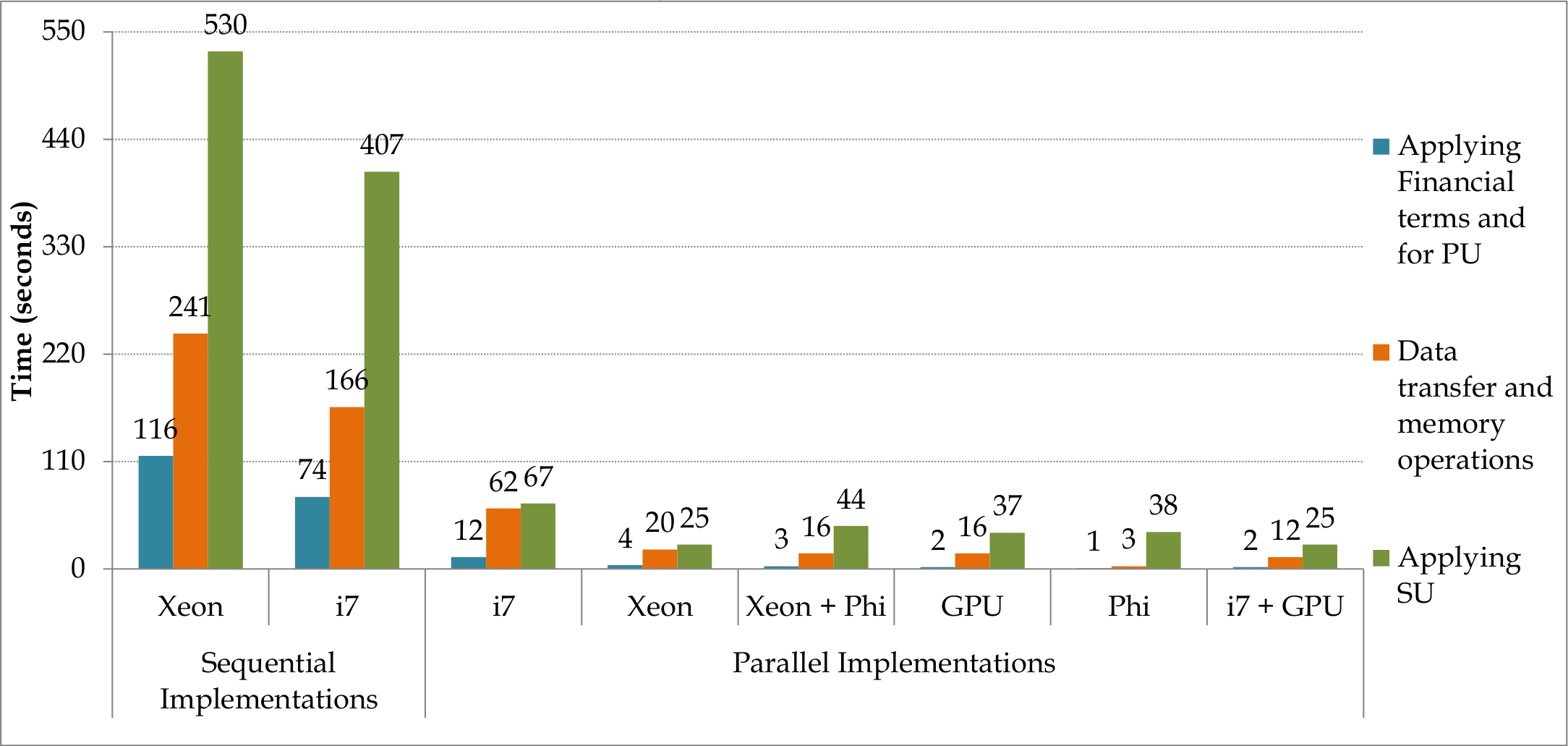}
	\caption{Time taken for fetching Events from memory and for look up of Loss Sets in memory, applying Financial Terms and performing other computations in PU, and applying SU on sequential and parallel implementations of ARA}
	\label{figure13}
\end{figure*}

Figure \ref{figure13} is a graph that summarises the key results from the experimental study. The set of three bars represents the time taken for (i) fetching Events from memory and for look up of Loss Sets in memory, (ii) applying Financial Terms and performing other computations in ARA, and (iii) applying secondary uncertainty on the sequential implementation on the CPU platforms and the parallel implementations on the multi-core CPU, many-core GPU and the Phi Co-processor. In each case, parameters specific to the implementation, such as the number of threads, were set to the best value identified during the experiments.

In the parallel implementations for ARA only considering PU, a speedup of 2.3x, of 20x and of 90x is achieved on the CPU, the GPU and the Phi respectively when compared to the corresponding sequential implementations (the parallel GPU results are compared against the sequential i7 results and the parallel Phi results are compared against the sequential Xeon results). 
A speedup of nearly 12x is achieved in the overall time for the implementation on the GPU and a speed up of 21x is obtained for the overall time for the implementation on the Phi in contrast to the respective sequential implementations on the CPU. 
For applying SU, multiple threading on the CPU is nearly between six to twenty two times faster than the sequential implementations. 
For the numeric computations on the GPU, an acceleration of approximately 26x is achieved over the sequential implementation.
Limited memory bandwidth is a bottleneck in the CPU that results in approximately 27\% and 53\% of the time spent for fetching Events and for look up of Loss Sets in memory for the sequential and parallel implementation on the CPU respectively. 
While the time for fetching Events and for look up of Loss Sets in memory have been significantly lowered on the GPU, 39\% of the total time is still used to this end. 
In the case of the Phi, this time is lowered to 3 seconds as against 16 seconds required by the GPU. 

On hybrid platforms, the combination of GPU and CPU outperform the Phi and CPU. 
The GPU and CPU perform ARA with PU and SU in only 41 seconds compared to 63 seconds required by Phi and CPU; the Phi and CPU is 50\% slower than the GPU and CPU platform, which is nearly 16x faster than the baseline implementation on i7. 
On individual accelerator platforms, without considering the time required for transferring data to the device from the host, the Phi only requires 42 seconds against 55 seconds required by the GPU; the Phi is 1.3 times faster than the GPU for ARA and 21x faster than the baseline implementation on Xeon. 

The Phi in the native mode seems faster than the combination of the Xeon and the Phi as a hybrid system. However, this is not always a practical solution since data needs to be copied offline to the memory of the Phi and is time consuming. In the case of ARA, copying data of 800,000 trials required more than three hours which is not included in the execution time. Although the execution takes only 42 seconds once all data resides in the Phi memory, it is less likely that most application can use the Phi in the native mode. The Phi is not intended to be used natively, but as a coprocessor. Hence, in the timing results the time taken for copying data to the Phi is not included.

In sequential or parallel implementations on CPU, accelerator and hybrid platforms require nearly 50\% or more time for applying SU in ARA. The majority of this time is required by the computations of the inverse CDF of the beta distribution. This calls for not only the development of fast methods to apply secondary uncertainty in risk analytics, but also the development of fast methods for the underlying statistical functions. Fast methods have been implemented for computing the inverse CDF of the symmetrical beta distribution \cite{invsymbetadist} which considers one shape parameter, but there are minimal implementations of fast assymetrical beta distribution that takes two shape parameters as required for the secondary uncertainty methodology reported in this paper.

\section{Conclusions}
\label{conclusions}
Hardware accelerators, such as GPUs and Phis, are attractive for accelerating applications that require HPC solutions, since they are low budget platforms. Risk analytics is one such domain that can hugely benefit from hardware accelerators. In this research, we set out to investigate which hardware accelerator is more beneficial for high-performance risk analytics. ARA, a Monte Carlo like simulation that considers primary and secondary uncertainties for estimating portfolio wide risk is chosen as a candidate application. Parallel algorithms to speedup the risk simulation are implemented on multi-core CPUs, hardware accelerators and hybrid platforms comprising CPUs and accelerators. Experimental studies are pursued on CPUs such as the Intel i7 and the Intel Xeon, and on accelerators, such as the NVIDIA GPU and the Intel Phi, for evaluating the performance of ARA. 

Both the Phi and the GPU are competing hardware accelerators and useful in different contexts for high-performance risk analytics. Overall, the experimental study indicates that for individual performance, when execution times are only considered without taking into account the data transfer times, the Phi outperforms the GPU by 23\% achieving a speed up of 21x over a baseline implementation; for a typical simulation of 800,000 trials with each trial comprising 1,000 events and 16 extended ELTs, the Phi individually requires only 42 seconds against 55 seconds required by the GPU. However, the Phi as an independent processor is impractical in a large number of cases and is best suited as a coprocessor in a hybrid platform. The Xeon and Phi hybrid platform is 50\% slower than the combination of i7 and the GPU. For the typical simulation on hybrid platforms, the i7 and GPU combination takes only 41 seconds, which is 16x faster than a baseline implementation, compared to 63 seconds required by Xeon and Phi.

\end{document}